\documentclass[usenatbib]{mnras}
\usepackage{pdflscape}
\usepackage{color}
\usepackage{IEEEtrantools} %
\usepackage{amsmath} %
\usepackage{amssymb} %
\usepackage{bm} %
\usepackage{upgreek} %
\usepackage[pdftex]{graphicx} %
\usepackage{multirow}
\usepackage{textcomp}
\usepackage[usenames,dvipsnames,svgnames,table]{xcolor}

\def\araa{ARA\&A}
\def\apj{ApJ}
\def\apjl{ApJ}
\def\apjs{ApJS}
\def\ao{Appl.~Opt.}

\def\aap{A\&A}
\def\aapr{A\&A Rev.}
\def\aaps{A\&AS}

\def\mnras{MNRAS}

\def\pra{Phys.~Rev.~A}

\def\pasa{Publications of the Astron. Soc. of Australia}

\def\pasj{PASJ}

\def\nat{Nature}

\newcommand{\sect}[1]{\text{Sect.~\ref{#1}}}
\newcommand{\fig}[1]{\text{Fig.~\ref{#1}}}
\newcommand{\tab}[1]{\text{Table~\ref{#1}}}
\newcommand{\app}[1]{\text{Appendix~\ref{#1}}}
\newcommand{\multitd}{\textsc{multi3d}}

\newcommand{\marcs}{\textsc{marcs}}
\newcommand{\stagger}{\textsc{stagger}}
\newcommand{\uppsala}{\textsc{uppsala}}

\newcommand{\mtd}{\textlangle3D\textrangle}
\newcommand{\FGK}{late-type}
\newcommand{\kms}{\mathrm{km\,s^{-1}}}

\newcommand{\teff}{T_{\mathrm{eff}}}
\newcommand{\lgg}{\log{g}}
\newcommand{\lggf}{\log{gf}}
\newcommand{\feh}{\mathrm{\left[Fe/H\right]}}

\newcommand{\xfe}[1]{\mathrm{\left[#1/Fe\right]}}
\newcommand{\lgeps}[1]{\log{\epsilon_{\mathrm{#1}}}}
\newcommand{\sh}{S_{\mathrm{H}}}
\newcommand{\lgt}{\log{\tau_{500}}}

\newcommand{\FeI}{\ion{Fe}{I}}
\newcommand{\FeII}{\ion{Fe}{II}}
\newcommand{\FeIII}{\ion{Fe}{III}}
\newcommand{\logepsdef}{$\lgeps{Fe}=\log_{10}\left(\frac{N_{\mathrm{Fe}}}{N_{\mathrm{H}}}\right)+12$}
\newcommand{\markaschanged}[1]{#1}
\title[Fe in late-type stars --- III]{Non-LTE line
formation of Fe in late-type stars --
III.~3D non-LTE analysis of metal-poor stars}
\author[A.~M.~Amarsi, K.~Lind, M.~Asplund, P.~S.~Barklem, 
R.~Collet]{A.~M.~Amarsi$^{1}$\thanks{E-mail: 
\href{mailto:anish.amarsi@anu.edu.au}{anish.amarsi@anu.edu.au}},
K.~Lind$^{2,3}$,
M.~Asplund$^{1}$,
P.~S.~Barklem$^{3}$,
and R.~Collet$^{4}$\\
$^{1}$ Research School of Astronomy and Astrophysics,
Australian National University, Canberra, ACT 2611, Australia\\
$^{2}$ Max Planck Institute f\"ur Astronomy, K\"onigstuhl 17, 
D-69117 Heidelberg, Germany\\
$^{3}$ Theoretical Astrophysics, Department of Physics and Astronomy, 
Uppsala University, Box 516, SE-751 20 Uppsala, Sweden \\
$^{4}$ Stellar Astrophysics Centre, Department of Physics and Astronomy, 
Aarhus University, Ny Munkegade 120, DK-8000 Aarhus C, Denmark}
\begin{document}

\date{Accepted ---. Received ---; in original form ---}
\pagerange{\pageref{firstpage}--\pageref{lastpage}} \pubyear{---}

\maketitle 
\label{firstpage}
\begin{abstract}
As one of the most important elements in astronomy,
iron abundance determinations need to be as accurate as possible.
We investigate the accuracy of spectroscopic iron abundance analyses
using archetypal metal-poor stars. 
We perform detailed 3D non-LTE radiative transfer calculations 
\markaschanged{based on} 3D hydrodynamic \stagger~model atmospheres,
and employ a new model atom that includes 
new quantum-mechanical neutral hydrogen
collisional rate coefficients.
With the exception of the red giant HD122563,
we find that the 3D non-LTE models achieve 
\FeI/\FeII~excitation and ionization balance 
as well as not having any trends with equivalent width
to within modelling uncertainties of $0.05\,\mathrm{dex}$, all 
without having to invoke any microturbulent broadening;
for HD122563 we predict that the current best
parallax-based surface gravity is overestimated by $0.5\,\mathrm{dex}$.
Using a 3D non-LTE analysis, 
we infer iron abundances from the 3D model atmospheres that are 
roughly $0.1\,\mathrm{dex}$~higher than 
corresponding abundances from 1D \marcs~model atmospheres;
these differences go in the same direction as the non-LTE effects
themselves. We make available grids of departure coefficients,
equivalent widths and abundance corrections,
calculated on 1D \marcs~model atmospheres and 
horizontally- and temporally-averaged 3D \stagger~model atmospheres.

\end{abstract}
\section{Introduction}
\label{introduction}

Iron is one of the most important elements in astronomy.
The high binding energy of $^{56}\text{Fe}$~makes it 
the end product of nucleosynthesis in
the cores of massive stars \citep[e.g.][]{2002RvMP...74.1015W},
as well as the main product of Type-Ia supernovae through
the radioactive decay of 
$^{56}\text{Ni}$~\citep[e.g.][]{1986ARA&amp;A..24..205W}.
Consequently, iron is highly abundant in the cosmos
\citep[{$\lgeps{Fe\odot}\approx7.47$\footnote{\logepsdef} in the
solar photosphere;}][]{2015A&amp;A...573A..26S}; 
this, coupled with its rich electronic structure,
leads to numerous iron spectral lines that are easy to detect,
even in extremely metal-poor stars 
\citep[e.g.][]{2013ApJ...762...26Y,2016ApJ...817...53S}.
Abundance ratio relations of the form
$\xfe{X}$~against $\feh$~are commonly used to interpret the
chemical evolution of galaxies 
\citep[e.g.][]{1979ApJ...229.1046T,
1993A&amp;A...275..101E,
2006ApJ...653.1145K}.
Numerous neutral iron lines with different strengths and
atomic properties can be used to constrain 
the effective temperature $\teff$~(excitation balance)~and
microturbulent broadening parameter $\xi$ needed in 1D
spectroscopic analyses,
while the presence of singly ionized lines also allows constraints 
on the surface gravity $\lgg$~\citep[ionization balance; 
e.g.][]{2012MNRAS.427...27B}.
Finally, as abundances and opacities are interlinked,
accurate iron abundances are important for stellar interior modelling:
enhanced iron opacities helped to explain the discrepancies
between theoretical and observed period ratios of Cepheid variables
\citep{1991ApJ...371L..73I,1992ApJ...385..685M}, and might
similarly lead to a resolution of the solar modelling problem
\citep[e.g.][]{2009ApJ...705L.123S,2015Natur.517...56B}.

However, in classical abundance analyses, 
the accuracy of iron abundance determinations 
is limited by the assumption
of Saha-Boltzmann energy partitioning,
or local thermodynamic equilibrium (LTE).
There is a rich literature on the non-LTE effects in iron
in \FGK~stars
\citep[e.g.][]{1971PASJ...23..217T,
1972ApJ...176..809A,
1985ASSL..114..231S,
1991A&amp;A...242..455T,
1999ApJ...521..753T,
2001A&amp;A...366..981G,
2001A&amp;A...380..645G,
2003A&amp;A...407..691K,
2005A&amp;A...442..643C,
2011A&amp;A...528A..87M,
2012MNRAS.427...27B,
2012MNRAS.427...50L,
2015ApJ...808..148S},
with a general consensus that neutral iron suffers
from overionization \citep[e.g.][Chapter 14]{2014tsa..book.....H}.
Qualitatively, non-local radiation originating from 
deep within the stellar atmosphere
drives an underpopulation of the lower levels of the minority 
\FeI~species with respect to LTE in the line-forming regions,
and hence a weakening of the \FeI~lines.
The non-LTE effects are enhanced by the
overexcitation of low-lying states ($\lesssim 2\,\mathrm{eV}$)~to
intermediate states (2 -- 5 eV), that are more susceptible to
overionization \citep{2012MNRAS.427...27B,2012MNRAS.427...50L}.
The populations of the majority \FeII~species are 
only slightly perturbed, and therefore non-LTE effects 
on \FeII~lines are usually insignificant.  

One of the main sources of uncertainty in non-LTE investigations 
stems from how neutral hydrogen collisions
are treated \citep[e.g.][]{2005ARA&amp;A..43..481A}.
In the absence of detailed quantum-mechanical calculations,
authors have typically either neglected such transitions, or have
resorted to the semi-empirical recipe of 
\citet{1968ZPhy..211..404D,1969ZPhy..225..483D}
as formulated by either \citet{1984A&amp;A...130..319S}
or \citet{1993PhST...47..186L}, and parametrized
with a collisional efficiency scaling factor $\sh$.
As discussed in \citet[][]{2011A&amp;A...530A..94B}~and
\citet{2016A&amp;ARv..24....9B},
both of these approaches can be prone to severe systematic errors.
Non-LTE investigations, in addition to
detailed quantum-mechanical calculations 
for the collisional cross-sections, are necessary
for quantifying these errors.

As well as the assumption of LTE, 
classical abundance analyses are characterized by
their use of theoretical one-dimensional (1D) hydrostatic model atmospheres.
Real model atmospheres are three-dimensional (3D) and dynamic, and the 
mean temperature stratification is influenced by 
the competing effects of
near-adiabatic cooling due to the expansion of ascending material 
and radiative heating due to the reabsorption of radiation
by spectral lines \citep[e.g.][and references therein]{2009LRSP....6....2N}.
Stationary 1D model atmospheres
in which radiative equilibrium is enforced
and convective effects are neglected are not
guaranteed to predict the correct atmospheric structure;
discrepancies with their 3D counterparts
are large at low metallicity,
where the temperature gradients in the 1D model atmospheres
are typically too shallow \citep{1999A&amp;A...346L..17A}. 

One-dimensional model atmospheres are also limited in their ability
to model the 
line strengthening, skewing, shifting, and
broadening effects by convective motions
\citep[e.g.][]{1980LNP...114..213N,2000A&amp;A...359..729A}.
The microturbulent and macroturbulent broadening parameters 
required in 1D analyses approximately account for only some of these effects;
they are essentially free parameters.
In contrast, convective motions are predicted by
ab initio three-dimensional (3D) hydrodynamic atmosphere simulations.
Consequently, spectral line formation calculations
using 3D radiative transfer and 3D model atmospheres
naturally include the aforementioned effects
without requiring any broadening parameters
\citep[e.g.][]{1982A&amp;A...107....1N,
1990A&amp;A...228..155N,
2000A&amp;A...359..729A}.

3D LTE investigations into iron line formation
have generally concluded that temperature and density inhomogeneities
and velocity gradients, 
as well as differences in the mean temperature stratification
significantly strengthen \FeI~lines with respect to 1D LTE,
particularly lower-excitation lines
\citep[e.g.][]{1980LNP...114..213N,
1999A&amp;A...346L..17A,
2007A&amp;A...469..687C,
2013A&amp;A...560A...8M,
2015A&amp;A...579A..94G}.
However, 3D effects and non-LTE effects on spectral line formation
are generally not orthogonal; hence an investigation into 
3D non-LTE iron line formation is warranted.
Hitherto, computational constraints have restricted 
such investigations to the Sun and to small model atoms
\citep[][]{2012A&amp;A...547A..46H,2013A&amp;A...558A..20H,
2015A&amp;A...582A.101H},
or have demanded the use of the
1.5D approximation \citep[where the columns of a 3D
model atmosphere are treated 
independently;][]{2001ApJ...550..970S,2005ApJ...618..939S},
while others are actively developing tools
for detailed 3D non-LTE radiative transfer 
\citep[][]{2014A&amp;A...566A..89H}.

We present detailed 3D non-LTE radiative transfer calculations
for iron in the metal-poor benchmark stars 
HD84937, HD122563, HD140283, and G64-12.
We employ 3D hydrodynamic \stagger~model atmospheres
\citep{2013A&amp;A...557A..26M},
and a new comprehensive model atom
that includes quantum-mechanical neutral hydrogen 
collisional rate coefficients~\citep{2016PhRvA..93d2705B}.
Thus, our modelling has removed the classical free parameters
that have hampered stellar spectroscopy for
many decades: mixing length parameters \citep{2015A&amp;A...573A..89M}, 
microturbulence and macroturbulence \citep{2000A&amp;A...359..729A,
2013A&amp;A...557A..26M},
Uns\"old enhancement \citep{2000A&amp;AS..142..467B},
and Drawin scaling factors \citep{2011A&amp;A...530A..94B,
2016A&amp;ARv..24....9B}.
We also present extensive grids of
departure coefficients, equivalent widths and
abundance corrections,
calculated on theoretical 1D hydrostatic \marcs~model 
atmospheres \citep{2008A&amp;A...486..951G}
and on horizontally- and temporally-averaged
3D \stagger~model atmospheres
\citep[henceforth {\mtd};][]{2013A&amp;A...560A...8M}.

Our results complement and extend those
reported in the
first paper in this series \citep{2012MNRAS.427...27B},
where a non-LTE investigation based on \mtd~model atmospheres
highlighted the importance of both the atmospheric structure
and of departures from LTE on Fe line formation
in metal-poor stars.
Our results are also an update on 
the non-LTE abundance corrections presented in the 
second paper in this series \citep{2012MNRAS.427...50L},
which were calculated prior to the completion
of the \stagger~grid of 3D model atmospheres \citep{2013A&amp;A...557A..26M},
and prior to the computation of quantum-mechanical neutral hydrogen 
collisional rate coefficients~\citep{2016PhRvA..93d2705B}.
The next paper in this series will report on
an investigation into the 3D non-LTE iron line formation
in the Sun and the solar iron abundance (Lind et al. in preparation).

The rest of this paper is structured as follows. 
We describe the methodology in \sect{method}.
This includes an overview of the 3D non-LTE code,
model atom, and model atmospheres,
\markaschanged{and a discussion} of the observations and the line lists.
We present and discuss our 3D non-LTE calculations
for the four metal-poor benchmark stars
in \sect{resultsbenchmark}.
We present our 1D and \mtd~grids of 
departure coefficients,
equivalent widths and abundance corrections 
in \sect{resultsgrids},
and explain how to access and use them.
We also discuss the general characteristics
of the 3D non-LTE effects in this section.
We summarize our results in \sect{conclusion}.

\section{Method}
\label{method}

\subsection{Code description}
\label{methodcode}

\subsubsection{Overview}
The main code used in this project is 
a customized version 
of \multitd~\citep{1999ASSL..240..379B,2009ASPC..415...87L}.
The code takes an iterative approach
to solving the statistical equilibrium equations
\citep[e.g.][Chapter 9]{2014tsa..book.....H},
following the multi-level Approximate Lambda Iteration (MALI)
preconditioning method of \citet{1992A&amp;A...262..209R}.
After each iteration, the code solves
the radiative transfer equation on short characteristics
using cubic-convolution interpolation of 
the upwind and downwind quantities,
and cubic Hermite spline interpolation 
for the source function
\citep{2013A&amp;A...549A.126I}.

Iron was assumed to be a trace element with
no feedback on the background (LTE) atmosphere
\citep[the so-called
restricted non-LTE problem;][]{1971ARA&amp;A...9..237H}.
This approximation was explored in
\citet[][]{2012MNRAS.427...50L};
the error incurred appears to be
much smaller than the non-LTE effects themselves. 

More information on \multitd~can be found in \citet{2016MNRAS.455.3735A}.
Below, we describe some of the 
changes and customizations that were made 
specifically for this project.

\subsubsection{Equation-of-state and background opacity}
\label{methodcodeeos}
A new equation-of-state (EOS) and opacity code
was written for this project.
This code was merged with \multitd~for the calculation
of all the background quantities,
replacing the \uppsala~opacity package 
that was originally for the \marcs~stellar atmosphere code
\citep{1975A&amp;A....42..407G,2008A&amp;A...486..951G}.
As with the \uppsala~opacity package,
the new code assumes LTE,
and the ionization potentials are reduced  
\citep[][Chapter 10]{1999spec.book.....T}
to account for departures from the ideal gas law
\citep[discussed in e.g.][]{1988ApJ...331..794H}.
As such, the two codes give consistent results 
provided that the input data are identical. 
Considerable effort was spent to bring
the atomic and molecular data up to date.
The continuous opacity sources were slightly updated compared
to those listed in \citet[][Appendix D]{2010A&amp;A...517A..49H};
one notable difference is that 
an analytical expression was used for the \ion{H}{I}~bound-free Gaunt factor,
and the tables of \citet{1988ApJ...327..477H}~were
used for the \ion{H}{I}~free-free Gaunt factor,
instead of the tables/graphs presented in \citet{1961ApJS....6..167K}.
Atomic and molecular line data were taken from 
the Kurucz online database
\citep{1995ASPC...78..205K}\footnote{\url{http://kurucz.harvard.edu/}}.
Partition functions were taken from~\citet{2016A&amp;A...588A..96B}.

LTE populations
and background continuous opacities were computed at runtime
at each gridpoint in the model atmosphere,
and the background source function was calculated analytically
by assuming all background processes to be thermalizing
\citep[such that it is equivalent to the
Planck function, {$S_{\nu}^{\text{bg}}=B_{\nu}$}; e.g.][Chapter
4]{2014tsa..book.....H}.
Background line opacities were precomputed
on a grid of temperature, density and wavelengths,
for a given chemical composition and 
assuming a microturbulent broadening parameter $\xi=1.0\,\kms$,
and were interpolated on to the model atmosphere at runtime.
The interpolation errors incurred are 
permissible given the inherent approximations 
and uncertainties involved, and amount to 
errors in the final inferred abundance 
that are of the order 0.01~dex in HD122563, and 
that are negligible in the other three stars.

After the populations had converged,
the final emergent Fe line spectra were
computed without the inclusion of any background line opacities,
and used directly for iron abundance determination.
This is justified because all of the
Fe lines used in the analysis are, by selection, 
free of blends (\sect{methodobservations}).

\subsubsection{Angle quadrature}
The adopted angle quadratures were selected
to obtain accurate results at minimal computational cost.
For the statistical equilibrium calculations in 3D,
4-point Lobatto quadrature for the integral over $\mu=\cos\theta$ 
on the interval [-1,1], and, for the non-vertical rays,
equidistant 4-point trapezoidal integration 
for the integral over $\phi$ 
on the interval [0,2$\uppi$] were adopted;
this equates to 10 rays over the unit sphere in total.
For the statistical
equilibrium calculations in 1D,
10-point Gaussian quadrature for the integral over $\mu$ 
on the interval [-1,1] was adopted.
The final emergent fluxes were calculated  
using 5-point Lobatto quadrature for the integral over $\mu$ 
on the interval [0,1], and, in the 3D case, 
equidistant 4-point trapezoidal integration 
for the integral over $\phi$
on the interval [0,2$\uppi$].
\citep[For more information on these angle quadratures
see, for example,][Chapter 8.]{1956itna.book.....H}

\subsubsection{Frequency parallelization}
Owing to the rich electronic structure of iron,
a large model atom, of the order 300 levels
and 5000 lines, is required
to realistically model the non-LTE effects
\citep[][]{1991A&amp;A...242..455T}.
As such, 3D non-LTE calculations are far more computationally
demanding than analogous calculations for less complex atoms
such as lithium \citep[e.g.][]{2003A&amp;A...399L..31A,
2013A&amp;A...554A..96L,2016A&amp;A...586A.156K} 
or oxygen \citep[e.g.][]{2009A&amp;A...508.1403P,
2015A&amp;A...583A..57S,
2015MNRAS.454L..11A,2016MNRAS.455.3735A},
which require model atoms of the order
20 to 25 levels and 50 to 100 lines.
\markaschanged{To proceed,}
the 3D domain-decomposed parallelization
scheme of \multitd~\citep[][]{2009ASPC..415...87L}~was extended
such that the radiative transfer equations at different frequencies
are also solved in parallel.
\markaschanged{Parallelization over frequency space is required
when performing detailed 3D non-LTE radiative transfer calculations
for complex atomic species in a practical amount of time.}

\subsubsection{Loss of significance issues}
A recurring numerical issue in our non-LTE computations is that of 
loss of significance when reducing the statistical equilibrium equations.
To illustrate how the problem may arise, 
consider two levels $i$~and $j$~that
are in relative LTE (i.e.~that satisfy the Saha-Boltzmann equations).
For these levels, the quantity
\phantomsection\begin{IEEEeqnarray}{rCl}
    \delta&=&n_{i} C_{i\,j}-n_{j} C_{j\,i}\,,
\end{IEEEeqnarray}
where $C_{i\,j}$~is the collisional transition rate from level $i$~to
level $j$, should be identically zero.
Inside a computer such cancellations can depart considerably from zero 
when the collisional rates become large due to
finite numerical precision;
this is prone to occur for levels that 
are very close in energy.

In an earlier work \citep{2016MNRAS.455.3735A}~this problem was
overcome by switching from 64-bit double precision 
to 128-bit quadruple precision
to represent the relevant quantities.
In this work, to save time and memory 
double precision was used, and large collisional rates
were capped at some maximum value, $C_{i\,j}^{\text{cap}}$.
This prescription was found to give equivalent results to
those obtained by instead using a higher precision representation.

\subsection{Model atom}
\label{methodatom}


The energies and statistical weights of the 
states of iron, as well as the 
radiative and collisional transition rate coefficients
that connect them, are encapsulated in the model atom.
\markaschanged{The first three ionization stages
of iron were considered. The energies were all taken
from the Kurucz online database 
\citep{1995ASPC...78..205K}\footnote{\url{http://kurucz.harvard.edu/atoms/2600}}\footnote{\url{http://kurucz.harvard.edu/atoms/2601}},
which includes laboratory values for observed levels
\citep[the values of which are same as those found in
the NIST atomic spectra database;][]{NIST_ASD} \footnote{\url{http://www.nist.gov/pml/data/asd.cfm}}.
The oscillator strengths were also taken from
the Kurucz online database,
but were replaced with laboratory values where
the latter were available;
in particular, we list the sources for 
the oscillator strengths used in the
subsequent abundance analysis in \app{linelist}.
Photoionization cross-sections were taken from the Iron Project 
\citep[][and private communication]{1997A&amp;AS..122..167B}.}

\markaschanged{Collisional transitions are responsible
for thermalizing the system and bringing the results
closer to LTE. 
The rates of excitation by electron collisions
were calculated using the formula of
\citet[][Sect.~3.6.2]{2000asqu.book.....C}~for
permitted transitions, 
based on the 
Born and Bethe approximations and a semi-empirical Gaunt factor
\citep{1962ApJ...136..906V}.
The same formula was adopted for forbidden transitions,
using an effective oscillator strength of $f_{\text{forb.}}=0.005$.
The rates of \FeI~ionization by electron collisions
were calculated using the formula of
\citet[][Sect.~3.6.1]{2000asqu.book.....C},
based on an empirical correction 
to the Bethe cross-section \citep{percival1966cross},
and the rates of \FeII~ionization by electron collisions
were taken from the CHIANTI atomic database
\citep{1997A&amp;AS..125..149D,2013ApJ...763...86L}
\footnote{\url{http://www.chiantidatabase.org}}.}

Towards lower metallicities the electron number density 
diminishes and neutral hydrogen atoms become
increasingly more important thermalizing agents. 
The often used semi-empirical recipe of
\citet{1968ZPhy..211..404D,1969ZPhy..225..483D},
as formulated by either \citet{1984A&amp;A...130..319S}~or 
\citet{1993PhST...47..186L},
is based on
the classical electron-ionization cross-section
of \citet{thomson1912xlii}~and 
does not reflect the actual physics of the 
collisional interactions \citep{2011A&amp;A...530A..94B,
2016A&amp;ARv..24....9B}.
Although large errors in the spectrum modelling may be rectified by using 
a global collisional efficiency 
scaling factor $\sh$~\citep[provided
that it can be calibrated reliably; e.g.][]{2005ARA&amp;A..43..481A},
such an approach is not ideal 
when there are significant relative errors in the rates  
(in which case multiple $\sh$~factors would be required).

We stress here that our model atom adopts new Fe+H collision rates 
calculated with the asymptotic two-electron method,
which was presented in \citet{2016PhRvA..93d2705B}~where it was
applied to Ca+H.
The calculation used here 
\markaschanged{is the same as that
which will be adopted in \citet{nordlanderkeller}: it}
includes 138 states of \FeI
and 11 cores of \FeII, leading to the consideration of 
17 symmetries of the FeH molecule
The data include both \FeI~collisional excitation, and 
charge transfer processes leading to Fe$^+$ + H$^-$.
\markaschanged{While \FeII~collisional excitation
is still described by the old recipe of 
\citet{1968ZPhy..211..404D,1969ZPhy..225..483D},
these reactions
have little influence on the overall results (Lind et al. in preparation).}
Thus, for the first time we perform non-LTE line formation calculations 
for iron using realistic quantum-mechanical 
neutral hydrogen collision data and
without needing to invoke $\sh$~or other largely 
influential free parameters.

\markaschanged{For practical reasons
it was necessary to reduce the complexity
of the model atom before proceeding 
with the 3D non-LTE calculations.
We defer a detailed discussion on how the model atom was reduced
to the next paper in this series (Lind et al. in preparation).
Briefly, all fine-structure levels were 
collapsed into terms, and 
high-excitation levels were merged into
super levels \citep[e.g.][Chapter 18]{2014tsa..book.....H},
using the mean energies of those
levels weighted by their statistical weights.
The effective oscillator strengths
of lines between collapsed levels
or super levels were obtained 
as the mean oscillator strength weighted
by the lower-level statistical weight and
the wavelength \citep[][Equation 1]{2012MNRAS.427...27B}.
Furthermore, radiative transitions that had an insignificant effect
on the overall statistical equilibrium
as determined by tests on various 1D and \mtd~model atmospheres,
were removed from the model atom.
Tests on 1D and \mtd~model atmospheres suggest the error
incurred in the equivalent widths
from reducing the model atom is
less than $0.01\,\mathrm{dex}$.}

\markaschanged{The final model atom that was used in the 
statistical equilibrium calculations}
contains 421 \FeI~levels, 41 \FeII~levels, and the ground state
of \FeIII, with 38705~frequency points spread across
4000 \FeI~lines and 48 \FeI-\FeII~continua.
When determining the emergent intensities
after the populations had converged,
a second model atom was used,
that contains 838 \FeI~levels, 116 \FeII~levels,
and the ground level of \FeIII;
the exact number 
of \FeI~and \FeII~lines
varied depending on the star being analysed.
The levels in the second model atom correspond to those in 
the first model atom, but with fine structure splitting
taken into account; 
the converged populations were redistributed 
under the assumption that collisional transitions between 
fine structure levels dominate over radiative transitions,
which is a sufficient condition for
them to have identical departure coefficients.

\subsection{Model atmospheres}
\label{methodatmosphere}

\begin{figure*}
\begin{center}
\includegraphics[scale=0.65]{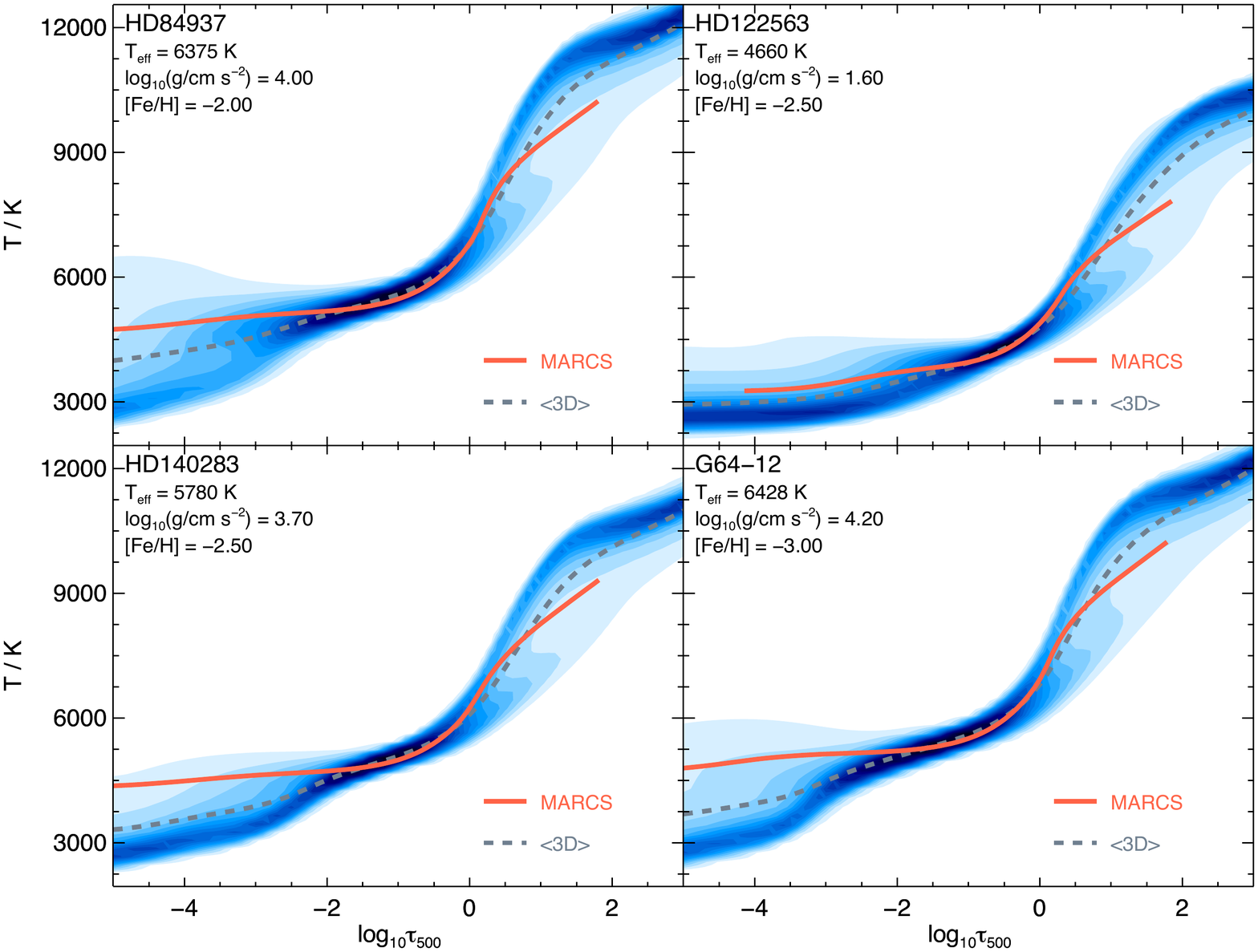}
\caption{Probability distributions of the gas temperature 
in 3D \stagger~model atmospheres of the benchmark stars
on surfaces of equal vertical $\lgt$.
Also shown are corresponding \marcs~and \mtd~temperature-$\lgt$~relations.
The \marcs~model atmospheres shown were found by interpolation
on to the stellar parameters listed in \tab{benchmark}.
The \mtd~model atmospheres shown
correspond to the horizontally- and temporally-averaged 3D \stagger~model 
atmospheres themselves, on surfaces of equal vertical $\lgt$.}
\label{model}
\end{center}
\end{figure*}

\subsubsection{3D hydrodynamic models}
\label{methodatmosphere3d}
\begin{table*}
\begin{center}
\caption{Literature values for the
stellar parameters of the four
metal-poor benchmark stars analysed in this work.
Also shown are the stellar parameters
of the corresponding 3D \stagger~model atmospheres. 
of the benchmark stars \citep{2011JPhCS.328a2003C}.}
\label{benchmark}
\begin{tabular}{l c c  c c  c c}
\hline
\multirow{2}{*}{Star} &
\multicolumn{2}{c}{$\teff/\mathrm{K}$} &
\multicolumn{2}{c}{$\log_{10}\left(g/\mathrm{cm\,s^{-2}}\right)$} &
\multicolumn{2}{c}{$\left[\mathrm{Fe/H}\right]$} \\ 

& 
Measured & Model &
Measured & Model &
Measured & Model \\
\hline
\hline

HD84937 &
$6356\pm97^{\text{a}}$ & $6375\pm21$ & 
$4.06\pm0.04^{\text{b}}$ &  $4.0$ &
$-2.03\pm0.08^{\text{e}}$ & $-2.0$ \\

HD122563 &
$4587\pm60^{\text{b}}$ & $4660\pm8$ &
$1.61\pm0.07^{\text{b}}$ & $1.6$ &
$-2.57\pm0.24^{\text{e}}$ & $-2.5$ \\

HD140283 &
$5591\pm162^{\text{c}}$ & $5780\pm10$ &
$3.65\pm0.06^{\text{c}}$ & $3.7$ &
$-2.40\pm0.07^{\text{e}}$ & $-2.5$ \\

G64-12 &
$6435\pm150^{\text{d}}$ & $6428\pm13$ &
$4.26\pm0.15^{\text{d}}$ & $4.2$ &
$-3.10\pm0.17^{\text{e}}$ & $-3.0$ \\

\hline
\hline
\end{tabular}
\end{center}
\medskip
a: Effective temperature based on
interferometric surface brightness relations calibrations 
of \citet{2004A&amp;A...426..297K},
from \citet{2015A&amp;A...582A..49H};
b: Effective temperatures based on interferometric measurements
and surface gravities based on parallax measurements
of \citet{2014ApJ...792..110V},
from \citet{2015A&amp;A...582A..49H};
c: Effective temperature based on interferometic measurement, 
accounting for the uncertainty in reddening,
and surface gravity based on parallax measurement of \cite{2013ApJ...765L..12B},
from \cite{2015A&amp;A...575A..26C};
d: Effective temperature based on H$\beta$~and
surface gravity based on Str\"omgren photometry
from \citet{2007A&amp;A...469..319N},
with our own estimates of their uncertainties;
e: Metallicities based on \mtd~non-LTE spectroscopy
from \citet{2012MNRAS.427...27B}.
\end{table*}

The 3D hydrodynamic model atmospheres were
computed using the
\stagger~code~\citep{nordlund19953d,1998ApJ...499..914S,
2011JPhCS.328a2003C,2013A&amp;A...557A..26M}.
The simulations are of the box-in-a-star variety
with Cartesian geometry.
\markaschanged{The model atmospheres were first presented in
\citet[][]{2011JPhCS.328a2003C}
and were used (after averaging in space and time)
by \citet{2012MNRAS.427...27B};
the model atmospheres have been updated since then
to better match the observed stellar parameters.}
We list the stellar parameters of
the model atmospheres in \tab{benchmark},
alongside recent measurements of the
actual stellar parameters of the corresponding benchmark stars.
The interferometric effective temperatures 
and parallax-based
surface gravities have little dependence on model atmospheres and
spectroscopy. 
We show the distribution of gas temperature
on surfaces of equal $\lgt$~in \fig{model},
alongside the corresponding \mtd~and 1D model 
atmospheres (\sect{methodatmosphere1d}).

The 3D models of HD84937, HD140283 and G64-12 have
mesh sizes $240\times240\times240$~points;
the model of HD122563 was recently recomputed with higher resolution,
having a mesh size $480\times480\times240$~points, 
where the last dimension represents the vertical.
For the 3D non-LTE radiative transfer calculations 
the horizontal meshsize was
reduced to $60\times60$~points by selecting every fourth point
(HD84937, HD140283, and G64-12) or every eighth point
(HD122563) in the $x$-~and $y$-directions.
The vertical meshsize was reduced to $101$~points.
This was done in two steps:
first, layers for which the vertical optical
depth at 500nm satisfied $\lgt>3$, in every column, were removed.
Second, the layers were interpolated 
such that the mean step in $\lgt$ was roughly constant
between layers.

Full 3D non-LTE radiative transfer calculations
were performed on three snapshots for each of the benchmark stars.
For each snapshot the EOS was determined 
for a range of values of $\lgeps{Fe}$, in steps of 0.5 dex.
Consistency between this value of $\lgeps{Fe}$~and
the EOS and background opacities was enforced.
However, consistency with 
the metallicity of the background atmosphere 
was not enforced, under the assumption 
that iron is a trace element with no feedback on the atmosphere.
This is obviously not strictly correct even for
these metal-poor stars, but is necessary
to make the calculations tractable; we intend to return to
this issue in a future work.

Computational constraints meant that
calculations were done on a few select 3D model atmospheres
rather than on an extended grid. 
In order to interpolate the results on to the
measured stellar parameters listed in \tab{benchmark},
the variations in the equivalent width with 
effective temperature and surface gravity
for a given line and 
iron abundance $\lgeps{Fe}$ were estimated 
using the grid of \marcs~model atmospheres (\sect{methodatmosphere1d}).

\subsubsection{1D hydrostatic models}
\label{methodatmosphere1d}
\begin{table}
\begin{center}
\caption{Extent of the grid of 1D \marcs~model atmospheres.
Consistency between the 
microturbulent broadening parameter $\xi$~used in the
non-LTE radiative transfer calculations,
and the microturbulence with which the model atmospheres were 
constructed, was not enforced. 
Missing models were obtained by interpolation,
extrapolation, or substitution, as described in the text.}
\label{onedgrid}
\begin{tabular}{l c c c}
\hline
Parameter & Min & Max & Step \\
\hline
\hline
$\teff/\mathrm{K}$ & $3500$ & $8000$ & $250$ \\
$\log_{10}\left(g/\mathrm{cm\,s^{-2}}\right)$ & $0.5^{\text{a}}$ & $5.0$ & $0.5$ \\
$\left[\mathrm{Fe/H}\right]$ & $-5.0$ & $0.5$ & $0.25$ \\
$\log_{10}\left(\xi/\kms\right)$ & $-0.1$ & $0.5$ & $0.3$ \\
\hline
\hline
\end{tabular}
\end{center}
\medskip
a: $\lgg\geq2.0$~for $\teff\geq5750\,\mathrm{K}$.
\end{table}

Radiative transfer calculations were also performed on
an extensive grid of 1D model atmospheres,
using both 1D hydrostatic \marcs~model 
atmospheres \citep{2008A&amp;A...486..951G}\footnote{\url{http://marcs.astro.uu.se/}}, 
and horizontally- and temporally-averaged 3D \stagger~model atmospheres
\citep[][]{2013A&amp;A...560A...8M}\footnote{\url{https://staggergrid.wordpress.com/mean-3d/}}, 
here denoted \mtd.
The standard-composition \marcs~models come in two varieties:
spherical and plane-parallel. 
For $\lgg\leq3.5$~the spherical models were used,
with microturbulence $2.0\,\kms$;
for greater values of $\lgg$~the plane-parallel models were used,
with microturbulence $1.0\,\kms$.
The \mtd~model atmospheres were obtained by
averaging the gas temperature and
logarithmic gas density from the 3D \stagger~model atmospheres 
on surfaces of equal time and $\lgt$.
All other thermodynamic variables were calculated
from these quantities via the EOS.

We state the extent of the grid in \tab{onedgrid}.
Not all of the models were available;
missing nodes that were within the boundaries of the grid 
were found by interpolation.
Where possible, in the regime $\feh\lesssim-3.0$,
missing model atmospheres beyond the boundaries of the grid
were substituted with
the model having the same effective temperature
and surface gravity,
and the closest value of $\feh$.
Missing \mtd~model atmospheres that could not be obtained by interpolation
or by the above procedure
were substituted with the corresponding \marcs~model atmospheres
in order to construct a complete grid.

To permit a fair comparison
between the results obtained with 
3D and \mtd~model atmospheres,
the \mtd~analysis of the benchmark stars was
based on calculations performed on
horizontally- and temporally-averaged
3D \stagger~model atmospheres themselves,
instead of on the calculations performed 
on the grid of \mtd~model atmospheres. 
Variations with stellar parameters were 
estimated in the same manner as was done
for the full 3D~analysis (\sect{methodatmosphere3d}).

\subsection{Observations}
\label{methodobservations}

HD84937, HD122563, and HD140283 spectra obtained with
UVES/VLT via the UVES Paranal Observatory Project 
\citep[POP;][]{2003Msngr.114...10B} were used;
the spectral resolving power is $R\approx80000$~and
the signal-to-noise ratio is typically around 300 to 500.
For G64-12 spectra obtained with
UVES/VLT taken on
2001-03-10~\citep{2004A&amp;A...414..931A} were used;
the spectral resolving power is $R\approx60000$~and
the signal-to-noise ratio is typically around 200 to 300.

\FeI~and \FeII~lines were selected for the analyses 
based on a number of criteria.
For maximum abundance sensitivity and diagnostic value,
the lines needed to be 
on the linear part of the curve of growth ($\log{W/\lambda}\lesssim-4.9$),
unblended, and detectable.
The lines needed to have 
quantum-mechanical van der Waals broadening data
\citep{1995MNRAS.276..859A,
1997MNRAS.290..102B,
1998MNRAS.296.1057B,
1998PASA...15..336B,
2005A&amp;A...435..373B},
and laboratory oscillator strengths.
To avoid complicating the analysis with
instrumental broadening and rotational broadening parameters 
\citep[as well as the macroturbulent 
broadening parameter, which is required in 
the case of 1D hydrostatic model atmospheres
to account for convective motions on
length scales larger than about one optical depth; 
e.g.][]{1992oasp.book.....G},
the analysis was based on a comparison of
equivalent widths, rather than the spectral line profiles themselves.
The equivalent widths were measured by direct integration,
which is the preferred method for very high signal-to-noise spectra
for which spectral lines may deviate from Gaussian- or Voigt-shaped profiles.
We list the selected \FeI~and \FeII~lines, 
and their measured equivalent widths, in \app{linelist}.

\subsection{Error analysis}
\label{methoderrors}

For a given effective temperature
and surface gravity, the weighted mean abundance is
\phantomsection\begin{IEEEeqnarray}{rCl}
\label{weightedmean}
    \textlangle\lgeps{Fe}\textrangle&=&\displaystyle\sum\limits_{i} 
    w_{i}\,{\lgeps{Fe}}_{i}\, ,
\end{IEEEeqnarray}
with the standard error in $n$~measurements
\phantomsection\begin{IEEEeqnarray}{rCl}
\label{standarderror}
    \sigma_{\text{SE}}^{2}&=&
    \frac{1}{n-1}\displaystyle\sum\limits_{i} w_{i}\,
        \left({\lgeps{Fe}}_{i}-\textlangle\lgeps{Fe}\textrangle\right)^2\, .
\end{IEEEeqnarray}
Here the normalized weights $w_{i}$~fold in
the stipulated errors in ${\lggf}_{i}$~and
the measurement errors in the observed equivalent widths $W_{i}$,
assuming these errors to be uncorrelated for individual lines $i$.
Errors in the inferred iron abundances ${\lgeps{Fe}}_{i}$~that
arise from errors in the effective temperatures and 
surface gravities are correlated. 
Therefore, these were folded into the overall error budget 
$\sigma$~in a separate step via 
$\sigma^{2}=\sigma_{\text{SE}}^{2}+\sigma_{\text{P}}^{2}$, where
\phantomsection\begin{IEEEeqnarray}{rCl}
\label{overallerror}
    \sigma_{\text{P}}^{2}=
    \left(\frac{\partial \textlangle\lgeps{Fe}\textrangle}
    {\partial \teff}\right)^{2}\sigma_{\teff}^{2}+
    \left(\frac{\partial \textlangle\lgeps{Fe}\textrangle}
    {\partial \lgg}\right)^{2}\sigma_{\lgg}^{2}\, .
\end{IEEEeqnarray}


\section{Metal-poor benchmark stars}
\label{resultsbenchmark}
\begin{table*}
\begin{center}
\caption{Inferred iron abundances of the benchmark stars using the best literature values of the effective temperatures and surface gravities (\tab{benchmark}). Absolute abundances $\lgeps{Fe}$~were converted to relative abundances $\feh$~by adopting for the solar iron abundance $\lgeps{Fe\odot}=7.47$, which is the value found by \citet{2015A&amp;A...573A..26S}~from an analysis based on a 3D \stagger~model atmospheres and \FeI~and \FeII~lines. The stipulated errors reflect the uncertainties in the equivalent widths, oscillator strengths, effective temperatures and surface gravities (\sect{methoderrors}). Other systematic modelling uncertainties have not been folded into the error analysis.}
\label{ionization}
\begin{tabular}{c c c c c c c c}
\hline
& \multirow{2}{*}{Star} & \multicolumn{2}{c}{1D} & \multicolumn{2}{c}{\mtd} & \multicolumn{2}{c}{3D} \\
& & $\feh^{\FeI}$ & $\feh^{\FeII}$ & $\feh^{\FeI}$ & $\feh^{\FeII}$ & $\feh^{\FeI}$ & $\feh^{\FeII}$ \\
\hline
\hline
\multirow{4}{*}{LTE}
& HD84937 & 
$-2.19\pm0.06$ &
$-2.07\pm0.02$ &
$-2.12\pm0.06$ &
$-2.00\pm0.02$ &
$-2.24\pm0.06$ &
$-2.00\pm0.02$ \\
& HD122563 & 
$-2.87\pm0.06$ &
$-2.51\pm0.04$ &
$-2.85\pm0.06$ &
$-2.43\pm0.05$ &
$-2.94\pm0.07$ &
$-2.43\pm0.04$ \\
& HD140283 & 
$-2.68\pm0.04$ &
$-2.41\pm0.05$ &
$-2.66\pm0.04$ &
$-2.30\pm0.04$ &
$-2.79\pm0.05$ &
$-2.30\pm0.04$ \\
& G64-12 & 
$-3.21\pm0.11$ &
$-3.17\pm0.11$ &
$-3.19\pm0.11$ &
$-3.09\pm0.10$ &
$-3.32\pm0.12$ &
$-3.11\pm0.11$ \\
\hline
\multirow{4}{*}{Non-LTE}
& HD84937 & 
$-2.02\pm0.06$ &
$-2.07\pm0.02$ &
$-1.98\pm0.06$ &
$-2.02\pm0.02$ &
$-1.90\pm0.06$ &
$-1.97\pm0.02$ \\
& HD122563 & 
$-2.78\pm0.07$ &
$-2.50\pm0.04$ &
$-2.77\pm0.07$ &
$-2.46\pm0.04$ &
$-2.70\pm0.07$ &
$-2.43\pm0.04$ \\
& HD140283 & 
$-2.49\pm0.04$ &
$-2.41\pm0.05$ &
$-2.45\pm0.04$ &
$-2.35\pm0.05$ &
$-2.34\pm0.05$ &
$-2.28\pm0.04$ \\
& G64-12 & 
$-2.98\pm0.11$ &
$-3.15\pm0.10$ &
$-2.95\pm0.11$ &
$-3.07\pm0.10$ &
$-2.87\pm0.11$ &
$-2.99\pm0.08$ \\
\hline
\hline
\end{tabular}
\end{center}
\end{table*}

\begin{figure*}
\begin{center}
\includegraphics[scale=0.65]{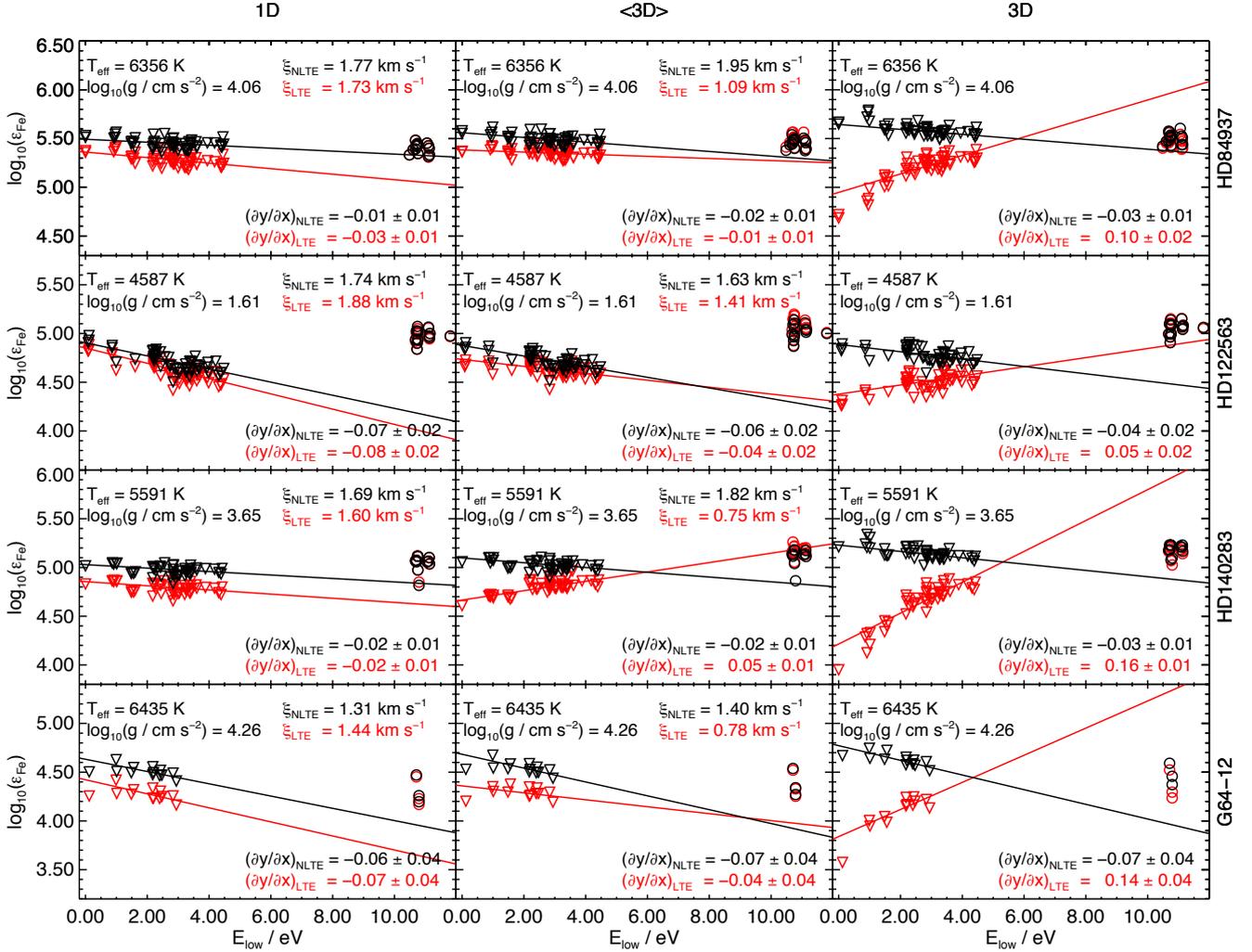}
\caption{Inferred iron abundance against excitation energy
for selected \FeI~and \FeII~lines.
\markaschanged{The excitation energies for both species
are given relative to the ground state of \FeI,
such that the ground state of \FeII~has
$E_{\mathrm{low}}=7.9024\,\mathrm{eV}$.}
Rows from top to bottom show the different benchmark stars:
HD84937, HD122563, HD140283, and G64-12.
Columns from left to right show the different paradigms:
1D radiative transfer with theoretical 1D~\marcs~model atmospheres,
1D radiative transfer with \mtd~\stagger~model atmospheres,
and full 3D radiative transfer with 3D \stagger~model atmospheres.
\markaschanged{\FeI~lines are indicated with 
black triangles (non-LTE) and red triangles (LTE);
\FeII~lines are indicated with 
black circles (non-LTE) and red circles (LTE).}
The least-squares trend with excitation energy of the \FeI~lines
is overdrawn; the standard error in the gradient
reflects the uncorrelated errors
arising from measurement errors in the 
observed equivalent widths as well as
correlated errors arising from
errors in the effective temperatures and
surface gravities (\sect{methoderrors}).}
\label{elo}
\end{center}
\end{figure*}

\begin{figure*}
\begin{center}
\includegraphics[scale=0.65]{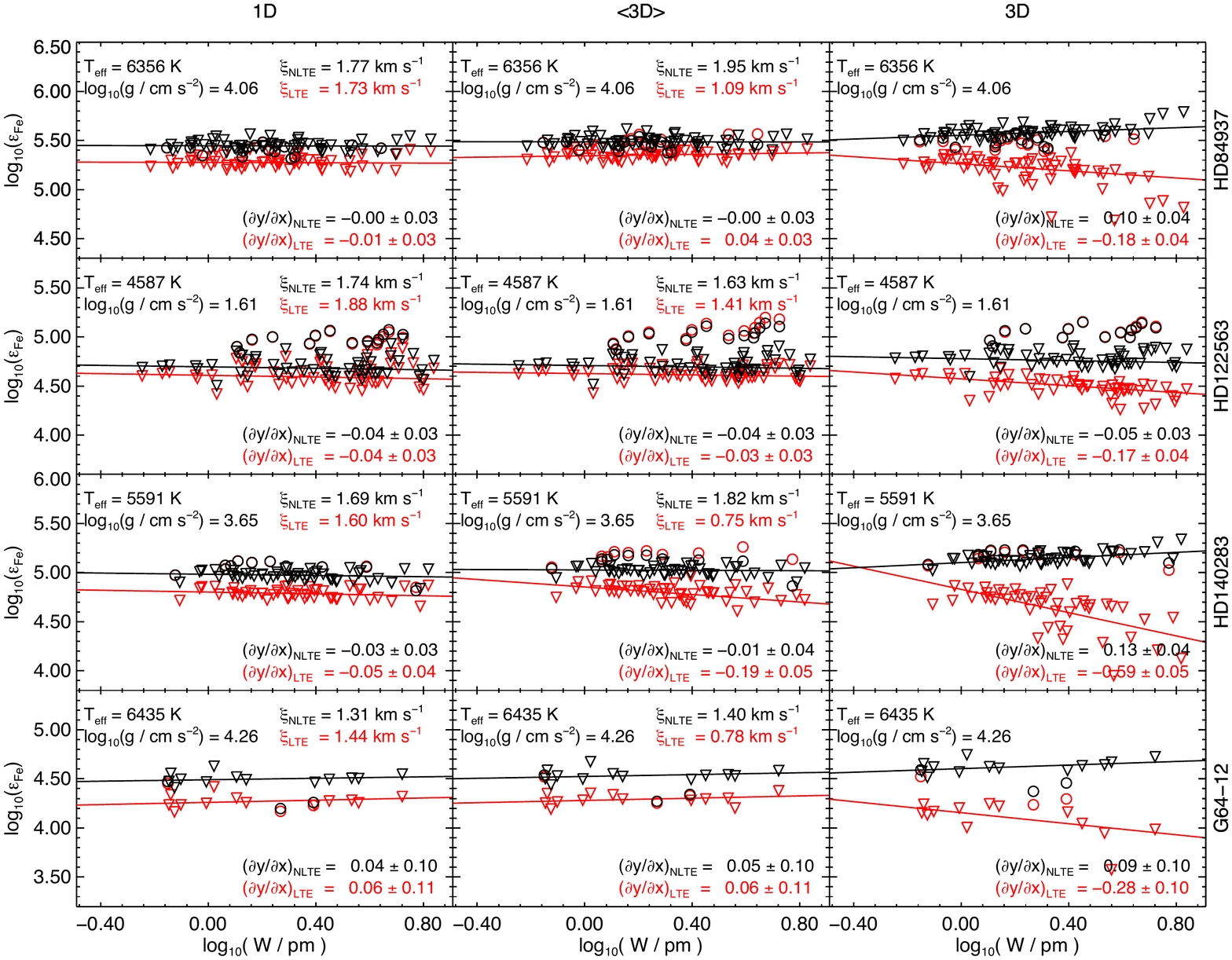}
\caption{Inferred iron abundance against equivalent width
for selected \FeI~and \FeII~lines.
Rows from top to bottom show the different benchmark stars:
HD84937, HD122563, HD140283, and G64-12.
Columns from left to right show the different paradigms:
1D radiative transfer with theoretical 1D~\marcs~model atmospheres,
1D radiative transfer with \mtd~\stagger~model atmospheres,
and full 3D radiative transfer with 3D \stagger~model atmospheres.
\markaschanged{\FeI~lines are indicated with 
black triangles (non-LTE) and red triangles (LTE);
\FeII~lines are indicated with 
black circles (non-LTE) and red circles (LTE).}
The least-squares trend with (logarithmic) equivalent width
of the \FeI~lines
is overdrawn; the standard error in the gradient
reflects the uncorrelated errors
arising from measurement errors in the 
observed equivalent widths as well as
correlated errors arising from
errors in the effective temperatures and
surface gravities (\sect{methoderrors}).
\markaschanged{The microturbulent parameter $\xi$~in
the 1D and \mtd~analyses was calibrated 
to flatten these trends, without considering 
the uncertainties in the effective temperatures
and surface gravities. 
If these uncertainties were considered
in the calibration of $\xi$,
then the trends in the first
two columns would be identically zero.}}
\label{lgeqw}
\end{center}
\end{figure*}

For a given star,
individual iron abundances were inferred for each line
by fitting the theoretical line strengths to the observed equivalent width.
The \markaschanged{1D analyses based on} \marcs~and \mtd~model atmospheres
contain a free parameter:
the microturbulent broadening parameter $\xi$. 
This was calibrated by removing the trend in the inferred
iron abundance from \FeI~lines against equivalent width.
We emphasize that in 3D non-LTE calculations
no microturbulence enters into the calculations
(\sect{methodatmosphere3d}),
which is one of the main advantages of using
3D model atmospheres \citep[e.g.][]{2000A&amp;A...359..729A}.

Assuming that the nominal measured effective temperatures and
surface gravities of the benchmark stars,
and their standard errors, are accurate
(\tab{benchmark}),
the reliability of the 3D non-LTE spectral line formation
calculations can be
assessed and quantified by considering the
difference between the iron abundances
that are a)~inferred from \FeI~lines
and from \FeII~lines (ionization balance);
b)~inferred from low-excitation \FeI~lines and
high-excitation \FeI~lines (excitation balance);
and c)~inferred from weak and strong Fe lines.
We illustrate this in \tab{ionization}, \fig{elo}~and \fig{lgeqw},
and discuss different aspects of the results below.

\subsection{3D non-LTE Fe line formation}
\label{resultsbenchmark3dnlte}

The trends in iron abundance 
with excitation energy 
inferred from \FeI~lines take the same shape in all cases,
having a slight, yet significant, negative gradient.
\fig{elo}~shows that
the mean difference between the iron abundances inferred from
the lowest and highest excitation \FeI~lines
is roughly $0.1\,\mathrm{dex}$~for HD84937 and HD140283,
which have the most significant gradients.
The corresponding trends in inferred iron abundance
with equivalent width are less significant,
with the trend for G64-12 being consistent with zero. 
\fig{lgeqw}~shows that,
as with the trend with excitation energy,
HD140283 has the most significant gradient:
the mean difference between the iron abundances inferred from
the weakest and strongest \FeI~lines
is of the order $0.1\,\mathrm{dex}$.

\tab{ionization}~shows that 
for HD84937, HD140283, and G64-12
the average 3D non-LTE iron abundances
inferred from \FeI~lines are larger than 
those inferred from \FeII~lines, by 
$0.07\,\mathrm{dex}$, $0.06\,\mathrm{dex}$, 
and $0.14\,\mathrm{dex}$~respectively. 
These discrepancies are consistent with the
$1\sigma$~standard errors in the mean for HD84937, HD140283,
and G64-12. 
For HD122563, on the other hand, the \FeI~result is smaller
than the \FeII~result, by $0.26\,\mathrm{dex}$,
a $3\sigma$~discrepancy. 
This large ionization-balance discrepancy
is not unique to this work
\citep[e.g.~][Sect.~4.5.4]{2012MNRAS.427...27B}.
Qualitatively, a similar discrepancy is found by the 
analyses based on the 1D model atmospheres, 
which suggests that the problem does not lie with
the 3D model atmosphere itself.
Rather, there may be something peculiar about 
some other aspect of the non-LTE calculations
in this region of parameter space,
such as the background opacities or the 
collisional rate coefficients, or 
about the stellar parameters that were adopted for this star.
Assuming the latter,
we predict that the surface gravity of this star is in fact
closer to $1.1\,\mathrm{dex}$.
With the upcoming Gaia Data Release 1
\citep{2015A&amp;A...574A.115M}, the parallax of 
HD122563 will be improved and thus our prediction tested.

Overall, with the exception of
the ionization-balance in the 
metal-poor red giant HD122563,
excitation-balance and ionization-balance
are obtained to a satisfactory level,
and the trends in inferred abundance with equivalent width
are flat. 
Small modelling errors of the order $0.05\,\mathrm{dex}$~are evident however.
That the trend with excitation energy takes the same shape
in all four stars, as well as in the 1D results (\sect{resultsbenchmark1d}),
suggests that the non-LTE effects in the 
low-excitation \FeI~may be slightly overestimated.
Slightly weaker non-LTE effects would also improve the ionization
balance in these stars (except for HD122563).
The use of a collisional efficiency scaling factor $\sh$~would
mask these modelling uncertainties,
as was possibly the case in our previous 1D work \citep{2012MNRAS.427...27B}.

\markaschanged{It is not obvious what exactly is causing 
this final shortcoming of a slight trend
in iron abundance with excitation potential; 
we note that in 1D analyses such trends
can normally be hidden by tuning the microturbulence parameter while in 
3D non-LTE there is no such luxury. 
The results could be hinting at inadequacies in the 
adopted formulae for collisional excitation and ionization by electron impact 
(\sect{methodatom}), which are based on
general approximations that do not take into account
the specific properties of the iron atom,
and are a clear weakness of the present modelling.
Then, an obvious next step in improving the non-LTE modelling
and in trying to resolve this issue
would be to calculate Fe+e$^{-}$~collisional cross-sections
using modern atomic structure and close-coupling methods 
such as R-matrix techniques
\citep[e.g.][]{1971JPhB....4..153B,1976AdAMP..11..143B,
1974CoPhC...8..149B,1978CoPhC..14..367B},
or B-spline R-matrix techniques
\citep[e.g.][]{2006CoPhC.174..273Z,2013JPhB...46k2001Z}.
Such data would at the very least rule out the collisional
modelling as the origin of the remaining discrepancies.}

\subsection{1D versus 3D non-LTE line formation}
\label{resultsbenchmark1d}

\tab{ionization} reveals that,
in non-LTE, larger iron abundances are inferred from 
\FeI~lines from the
3D model atmospheres than from the \mtd~model atmospheres.
This can be attributed to the  
effects of atmospheric inhomogeneities
and horizontal radiation transfer \citep{2013A&amp;A...558A..20H}.
Non-local radiation streaming in from surrounding regions
can strengthen and weaken \FeI~lines depending on the 
sign of the temperature contrast
\citep{2013A&amp;A...558A..20H},
in this case with a net line-weakening effect in 
the spatially-averaged \FeI~line profiles.
Slightly larger iron abundances are inferred from
the \mtd~model atmospheres than from the \marcs~model atmospheres.
This can be attributed to the differences
in the mean temperature stratification predicted
by 3D hydrodynamic models and by
1D hydrostatic models \citep{2012MNRAS.427...27B}.
Steeper temperature gradients in the outer layers 
of the \mtd~model atmospheres
lead to a greater amount of non-local radiation
and consequently a greater degree of overionizaton of 
neutral iron, which weakens the \FeI~lines
\citep{2012MNRAS.427...27B}.

\tab{ionization} also reveals that,
larger iron abundances are inferred from 
the \FeII~lines from the
3D model atmospheres than from the \mtd~model atmospheres,
and from \FeII~lines from the
\mtd~model atmospheres than from the \marcs~model atmospheres.
Since the \FeII~lines form close to LTE 
(\sect{resultsgridsnlte}), these observations must 
be attributed to differences in the local conditions
in the line-forming regions of the respective model atmospheres:
temperature and density inhomogeneities and velocity gradients, 
as well as differences in the mean temperature stratification.
The net line-weakening effects are consistent with what 
have been found in previous 3D LTE investigations,
for example~\citet{1999A&amp;A...346L..17A},
\citet{2007A&amp;A...469..687C}, 
\citet{2013A&amp;A...560A...8M}, and
\citet{2015A&amp;A...579A..94G}.

The microturbulent broadening parameter $\xi$, 
serves to improve the trends 
in inferred abundance with equivalent width
(\fig{lgeqw}), but also with excitation energy
(\fig{elo}),
in the 1D analyses relative to the 3D analyses,
this free parameter being fitted to remove the trend in 
inferred iron abundance from \FeI~lines 
with equivalent width.
Since $\xi$~affects strong lines, which tend to be of low-excitation,
more than it does weak (and high-excitation) lines,
it has a strong influence on these trends.
Its impact on the ionization balance is smaller,
which is why the ionization balance remains comparable 
in 1D non-LTE to that in 3D non-LTE
(\tab{ionization}).

HD122563~is best modelled in 3D, even without any free parameters.
\fig{elo}~shows that the trend in inferred
iron abundance with excitation energy
is significantly closer to zero 
with the 3D non-LTE analysis
than with the 1D non-LTE analysis 
using \marcs~and \mtd~model atmospheres.
\tab{ionization}~shows that 3D non-LTE analysis
results in the smallest discrepancy between the 
non-LTE iron abundances inferred from \FeI~lines
and those inferred from \FeII~lines.
This suggests that 3D models are necessary
for accurate spectroscopic analyses of HD122563,
and of similar types of stars;
however, the large ionization-balance
discrepancy found for this star (\sect{resultsbenchmark3dnlte})
prevents a stronger conclusion 
on this point from being drawn.

\subsection{LTE versus non-LTE line formation}
\label{resultsbenchmarklte}

Our results strongly favour 
non-LTE \FeI~line formation over LTE line formation.
\tab{ionization}~shows that the
discrepancy between the mean inferred abundances
from \FeI~and \FeII~lines is several standard errors
worse in 3D LTE than in 3D non-LTE.
Similarly, \fig{elo}~and \fig{lgeqw}
show that the trends in inferred iron abundance with
excitation energy are typically 
much steeper in 3D LTE than in 3D non-LTE,
by several standard errors.
This is not surprising:
steep temperature gradients and low temperatures
in the line-forming regions of low metallicity 3D models
make them prone to large departures
from LTE \citep{1999A&amp;A...346L..17A,2003A&amp;A...399L..31A}.

At first glance, the difference between the LTE results
and the non-LTE results
are less pronounced in the 1D calculations
than they are in the 3D calculations,
in particular with regards
to the trends in inferred iron abundance with
excitation energy and equivalent width.
This can be attributed to the free parameter $\xi$,
which serves to mask some of the short-comings of the assumption of LTE
in the 1D and \mtd~model atmospheres (\sect{resultsbenchmark1d}).
Also noteworthy are the hotter mean temperature
stratifications in the \marcs~model atmospheres
in the outer layers (compared to the mean temperature
stratification of the 3D model atmospheres; \fig{model}).
The theoretical low-excitation \FeI~lines are weaker
in higher temperature conditions in LTE,
meaning that a larger iron abundance is required to 
reproduce the observations.
This flattens the trend in inferred iron abundance
with excitation energy obtained with
the \marcs~model atmospheres in LTE.

In summary, it is particularly important to carry out
non-LTE calculations when using 3D model atmospheres at low metallicity
for elements susceptible to non-LTE effects.

\subsection{Best inferred iron abundances}
\label{resultsbenchmarkbestabund}
\begin{table}
\begin{center}
\caption{Best estimate of the iron abundances of the benchmark stars using the best literature values of the effective temperatures and surface gravities (\tab{benchmark}), and non-LTE results from the 3D \stagger~model atmospheres. Absolute abundances $\lgeps{Fe}$~were converted to relative abundances $\feh$~by adopting for the solar iron abundance $\lgeps{Fe\odot}=7.47$, which is the value found by \citet{2015A&amp;A...573A..26S}~from an analysis based on a 3D \stagger~model atmospheres and \FeI~and \FeII~lines. The stipulated errorsreflect the uncertainties in the equivalent widths, oscillator strengths, effective temperatures and surface gravities (\sect{methoderrors}). Systematic modelling errors have been estimated by taking half the difference between the abundances inferred from \FeI~and \FeII~lines.}
\label{bestabund}
\begin{tabular}{c c}
\hline
Star & $\feh$ \\
\hline
\hline
HD84937 & 
$-1.96\pm0.02\text{(stat)}\pm0.04\text{(sys)}$ \\
HD122563 & 
$-2.49\pm0.11\text{(stat)}\pm0.14\text{(sys)}$ \\
HD140283 & 
$-2.31\pm0.03\text{(stat)}\pm0.03\text{(sys)}$ \\
G64-12 & 
$-2.94\pm0.06\text{(stat)}\pm0.06\text{(sys)}$ \\
\hline
\hline
\end{tabular}
\end{center}
\end{table}


We provide our best inferred iron abundances for
the four benchmark stars in \tab{bestabund}.
These were computed from the mean
of the iron abundances 
inferred from the \FeI~and \FeII~lines using 
a 3D non-LTE analysis (i.e.~by combining
the last two columns and last four rows of 
\tab{ionization}), weighted by their
standard errors (and without systematic errors included).
Although the inferred abundances 
are consistent with those of
\citet{2012MNRAS.427...27B}, listed in \tab{benchmark},
to within the standard errors,
our results are typically higher than their results.
This can be attributed to 3D effects (\sect{resultsgrids3d})
as well as larger non-LTE effects
resulting from improved atomic data,
especially neutral hydrogen collisional 
rate coefficients (\sect{resultsgridsl12}).

\section{Grids of non-LTE abundance corrections} 
\label{resultsgrids}


\begin{figure*}
\begin{center}
\includegraphics[scale=0.6]{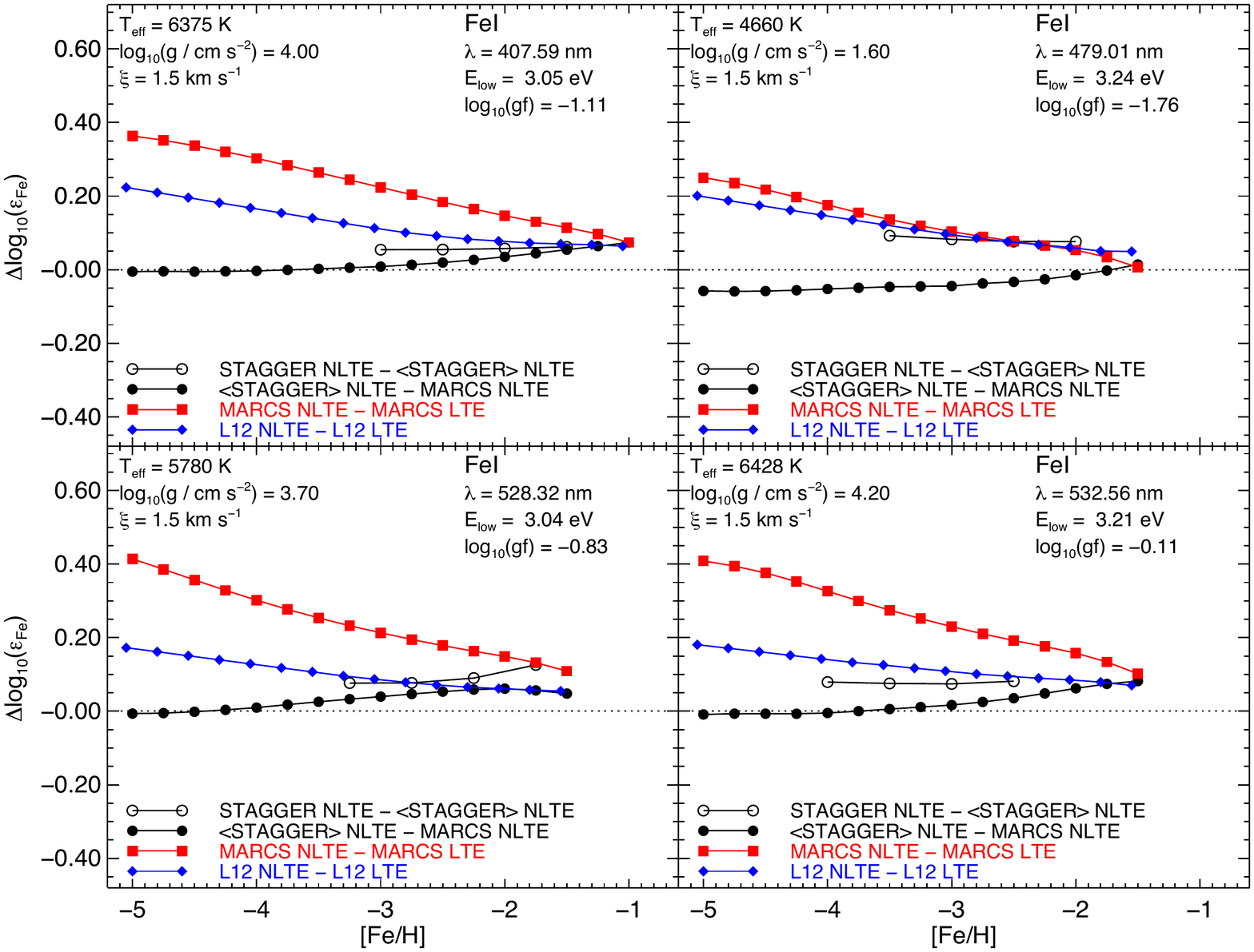}
\caption{Abundance corrections based on equivalent widths.
Each panel shows a typical high-excitation \FeI~line
for different sets of stellar parameters;
the plots are truncated after the line
becomes saturated in LTE
(reduced equivalent widths smaller than -4.9 dex).
Several different types of correction \markaschanged{are shown:}
3D non-LTE versus \mtd~non-LTE corrections;
\mtd~non-LTE versus 1D~non-LTE corrections;
1D non-LTE versus 1D LTE corrections from this work;
and 1D non-LTE versus 1D LTE abundance corrections, using 
\marcs~model atmospheres, from 
\citet{2012MNRAS.427...50L}.
The L12 results were shifted in $\feh$~to
put them on the same absolute scale,
where the reference solar iron abundance 
is $\lgeps{Fe\odot}=7.5$.}
\label{abcor0}
\end{center}
\end{figure*}

\begin{figure*}
\begin{center}
\includegraphics[scale=0.6]{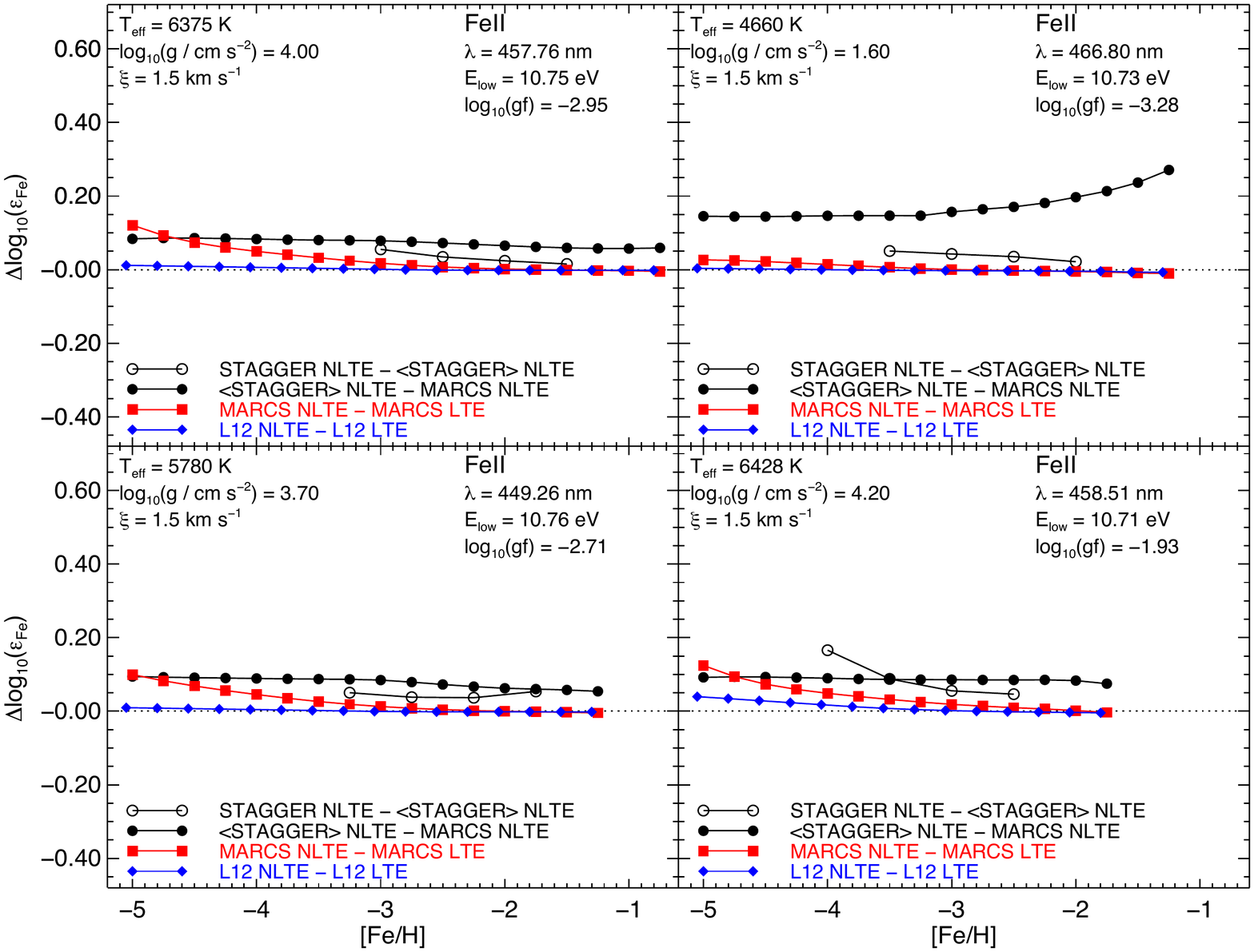}
\caption{Abundance corrections based on equivalent widths.
Each panel shows a typical \FeII~line
for different sets of stellar parameters;
the plots are truncated after the line
becomes saturated in LTE
(reduced equivalent widths smaller than -4.9 dex).
\markaschanged{The excitation energies
are given relative to the ground state of \FeI,
such that the ground state of \FeII~has
$E_{\mathrm{low}}=7.9024\,\mathrm{eV}$.}
Several different types of correction \markaschanged{are shown:}
3D non-LTE versus \mtd~non-LTE corrections;
\mtd~non-LTE versus 1D~non-LTE corrections;
1D non-LTE versus 1D LTE corrections from this work;
and 1D non-LTE versus 1D LTE abundance corrections, using 
\marcs~model atmospheres, from 
\citet{2012MNRAS.427...50L}.
The L12 results were shifted in $\feh$~to
put them on the same absolute scale,
where the reference solar iron abundance 
is $\lgeps{Fe\odot}=7.5$.}
\label{abcor1}
\end{center}
\end{figure*}

Grids of equivalent widths and abundance corrections
were constructed
for 2086 \FeI~lines and 115 \FeII~lines for which
accurate experimental atomic data exist.
These results can be accessed 
on the INSPECT database\footnote{\url{http://inspect-stars.com}},
or by contacting the authors.
Abundance corrections for other \FeI~and \FeII~lines
can be obtained by interpolation on the irregular grid
of line parameters
$\left(\lambda,\mathrm{E}_{\text{low}},\log g\,f\right)$;
interpolation routines for that purpose
are available upon request.

Grids of departure coefficients are also available,
which are more appropriate 
when higher precision and accuracy is required,
and when an analysis based on spectral line profile fitting,
rather than on equivalent widths, is called for. 
The departure coefficients can be used 
to correct the LTE populations in 
LTE spectral line synthesis codes,
to obtain non-LTE spectral lines and equivalent widths 
with almost no added computational cost. 

We illustrate abundance corrections,
based on equivalent widths, 
for typical \FeI~lines (\fig{abcor0})~and 
\FeII~lines (\fig{abcor1}),
and discuss their characteristics below.

\subsection{Non-LTE effect}
\label{resultsgridsnlte}

In line with previous non-LTE studies for iron
\citep[e.g.][]{1999ApJ...521..753T,
2003A&amp;A...407..691K,
2005A&amp;A...442..643C,
2011A&amp;A...528A..87M,
2012MNRAS.427...50L,
2015ApJ...808..148S},
\fig{abcor0}~shows that the 
\FeI~lines are significantly stronger in non-LTE than they are
in LTE, at least in the metal-poor regime.
The 1D non-LTE versus 1D LTE abundance corrections are positive, 
and increasing towards lower $\feh$.
This is consistent with the overionization picture
that we described in \sect{introduction}. 
The abundance corrections are in fact
typically monotonic with stellar parameters
increasing with increasing $\teff$, decreasing $\lgg$, and decreasing $\feh$
\citep[][Fig.~2]{2012MNRAS.427...50L}.

The overionization effect on \FeI~lines is very weak in
the solar-metallicity and metal-rich regimes (Lind et al. in preparation).
For some sets of stellar parameters,
the 1D non-LTE versus 1D LTE abundance corrections
are even slightly negative,
reaching $-0.05\,\mathrm{dex}$~at $\feh\approx0.5$.
Overionization is weaker in this regime;
photon suction driven by photon losses in the lines
becomes relatively important
\citep{2012MNRAS.427...27B}.
Since the absolute size of the corrections are so small
in this regime, 
the results are relatively sensitive to 
the amount of background opacity in the UV,
with more background opacity
serving to reduce the overionization effect
\citep[e.g.][Sect.~4.3]{1991A&amp;A...242..455T}.
We were therefore careful to use
recent opacity data for background lines and continua,
as we discussed in \sect{methodcodeeos}.

\fig{abcor1}~shows that the
non-LTE effects in \FeII~lines are not strictly
non-zero in the metal-poor regime. 
With \marcs~models,
they are negligible for $\feh\gtrsim-2.5$,
but grow with decreasing $\feh$.
Since 3D non-LTE abundances are significantly larger 
than 1D non-LTE abundances (\sect{resultsbenchmark1d}),
this suggests that the metallicities
of the most metal-poor stars 
\citep[][]{2002Natur.419..904C,2007ApJ...660L.117F,2015ApJ...806L..16B}
have hitherto been underestimated \citep[][]{nordlanderkeller}.
This is true regardless of whether (or which) \FeI~or 
\FeII~lines have been used.

\subsection{3D effect}
\label{resultsgrids3d}

The impact of using 3D radiative transfer
and 3D hydrodynamic model atmospheres on the inferred iron
abundance can be decoupled into two separate effects:
the effect of atmospheric inhomogeneities
and horizontal radiation transfer
\citep{2013A&amp;A...558A..20H},
and the effect of differences in the
mean temperature stratification.
In this section we compare the relative sizes of the
different effects, having discussed them 
already in \sect{resultsbenchmark1d}.
The former type of effects are quantified by the 
3D~versus \mtd~abundance corrections,
and the latter type are quantified by the 
\mtd~versus 1D~abundance corrections.
This decoupling is only to aid one's intuition:
the two effects are not orthogonal,
and their relative sizes are sensitive to
how the \mtd~model atmospheres are constructed.

For \FeI~lines the effect of atmospheric inhomogeneities
and horizontal radiation transfer
is small but significant:
\fig{abcor0}~shows that the 3D~non-LTE versus \mtd~non-LTE
abundance corrections are of the order $0.1\,\mathrm{dex}$.
This appears to be more severe than the effect 
of differences in the mean temperature stratification:
the magnitudes of the \mtd~non-LTE versus 1D non-LTE abundance corrections
for \FeI~lines are typically  
smaller than $0.05\,\mathrm{dex}$.
For \FeII~lines the two effects tend to be of comparable size
($0.05$-$0.10\,\mathrm{dex}$)~and are of the same
sign, compounding each other. 
This means that 3D effects are more pronounced 
in \FeII~lines than in
\FeI~lines, amounting to 3D non-LTE abundances
that are of the order $0.1\,\mathrm{dex}$~higher
than in 1D non-LTE with \marcs~model atmospheres.

In other words, with 1D \marcs~model atmospheres and
non-LTE radiative transfer, both \FeI~and \FeII~lines
underestimate the true iron abundance
by roughly $0.1\,\mathrm{dex}$~(as judged by 
the 3D non-LTE analysis; cf.~\sect{resultsbenchmark}, \tab{ionization}).
Obviously \FeI~lines in LTE suffer from
even larger systematic errors at low $\feh$~and 
thus their use should be avoided.

\subsection{Comparison with {\citet{2012MNRAS.427...50L}}}
\label{resultsgridsl12}

The main differences between the 
1D radiative transfer calculations
in 1D \marcs~model atmospheres
carried out in this work,
and the analogous calculations 
carried out by \citet{2012MNRAS.427...50L}, 
relate to the model atom and in particular to the treatment
of the neutral hydrogen collisional rate coefficients.
\citet{2012MNRAS.427...50L}~used semi-empirical hydrogen collisional
rate coefficients from
\citet{1968ZPhy..211..404D,1969ZPhy..225..483D}~with
collisional efficiency scaling factor $\sh=1.0$,
which was calibrated in \citet{2012MNRAS.427...27B}~so as to obtain
ionization balance in benchmark stars.
As we discussed in 
\sect{introduction}~\citep[see 
also][]{2011A&amp;A...530A..94B,2016A&amp;ARv..24....9B},
this empirical approach is based on a classical formula
that does not provide \markaschanged{a good description} of the actual
quantum-mechanical physics. 
In contrast, the model atom employed in this work utilized
new quantum-mechanical calculations 
for the neutral hydrogen collisional rate coefficients,
which are expected to be more accurate
than the previous classical description
by orders of magnitude \citep{2016PhRvA..93d2705B}.

The 1D non-LTE versus 1D LTE abundance corrections
predicted by \citet{2012MNRAS.427...50L}~for \FeI~and \FeII~lines
using \marcs~model atmospheres
are typically smaller than the corresponding
abundance corrections obtained here
by about 0.1 to 0.2~dex; for metal-poor giants
the two sets of calculations are more similar.
A smaller choice of $\sh$~in their calculations
would have increased their non-LTE effects,
thereby improving the agreement between their results and our own. 
It is important to appreciate, however, that calibrating $\sh$~masks
inadequacies in the models, that are not limited to
the description of the neutral hydrogen collisional
rate coefficients themselves.

\section{Conclusions}
\label{conclusion}

We have presented the first 3D non-LTE line formation calculations
for iron with quantum mechanical neutral hydrogen
collisional rate coefficients data.
The calculations were done for four metal-poor
benchmark stars: HD84937, HD122563, HD140283, and G64-12.
This type of analysis benefits from the absence of two free parameters:
the microturbulent broadening parameter $\xi$,
necessary in analyses based on 1D model atmospheres to
account for line broadening by convective motions,
and the collisional efficiency scaling factor $\sh$,
necessary in analyses based on the classical semi-empirical recipe of 
\citet{1968ZPhy..211..404D,1969ZPhy..225..483D}~for the
neutral hydrogen collisional rate coefficients. 

Our main conclusions are:
\begin{itemize}
\item{With the exception of the metal-poor red giant 
HD122563, the 3D non-LTE spectral line formation calculations
appear to be accurate to about
$\pm0.05\,\mathrm{dex}$ in inferred iron abundance.
Systematic modelling errors of this order 
are possibly due to the non-LTE effects being 
overestimated, which has hitherto gone
unnoticed because they were hidden 
in the calibration of the free parameter $\sh$.
\markaschanged{Modern quantum-mechanical 
electron collision cross-sections for iron
are needed to place the collisional modelling on a firm footing.
It is extremely important that such calculations
be done and we advocate that this should be investigated 
as the possible source of remaining discrepancies 
before considering more exotic explanations.}}
\item{We predict that the parallax-based surface gravity
of HD122563 is overestimated, on the basis that,
in contrast to the other three stars,
much larger non-LTE effects than what we currently
predict would be required
to rectify the observed ionization-imbalance.
We predict $\lgg\approx1.1$~for this star.}

\item{The new quantum-mechanical neutral hydrogen
collisional rate coefficients
result in larger non-LTE effects
than do the recipe of
\citet{1968ZPhy..211..404D,1969ZPhy..225..483D}~with $\sh=1.0$.}

\item{3D effects in \FeI~lines caused by atmospheric
inhomogeneities and horizontal radiation transfer
are typically more severe than 
those caused by differences
in the mean temperature stratifications;
the two effects in \FeII~lines are typically of
comparable size and have the same sign.
Overall, the 3D effects 
go in the same direction as the non-LTE effects,
requiring positive abundance corrections.
\FeII~lines are slightly more prone to these effects than \FeI~lines.}

\item{As a consequence of the 3D effects~and 
larger non-LTE effects, our 
inferred iron abundances for the four benchmark stars are 
slightly higher than in our previous 1D non-LTE
study \citep{2012MNRAS.427...27B}.}
\end{itemize}

Extended grids of abundance corrections and
equivalent widths based on full 3D non-LTE calculations 
are still somewhat beyond reach with today's
computational resources.
Nevertheless, we have made available 
grids of departure coefficients and abundance corrections 
based on 1D \marcs~model atmospheres
and \mtd~model atmospheres (\sect{resultsgrids}).
A future work may consider the construction of 
3D LTE grids of abundance corrections for \FeII~lines;
such grids would be valid in the regime $\feh\gtrsim-2.5$.
We recommend that such grids be implemented into
future spectroscopic stellar analyses
that are based on \FeI~and \FeII~lines,
including large-scale surveys such as
APOGEE~\citep{2015arXiv150905420M},
GALAH~\citep{2015MNRAS.449.2604D},
4MOST~\citep{2014SPIE.9147E..0MD},
and WEAVE~\citep{2014SPIE.9147E..0LD}.

\section*{Acknowledgements}
\label{acknowledgements}
AMA and MA are supported by the Australian
Research Council (ARC) grant FL110100012.
Data from the UVES Paranal Observatory Project 
(ESO DDT Program ID 266.D-5655) was used in this work.
KL acknowledges funds from the Alexander von Humboldt Foundation 
in the framework of the Sofja Kovalevskaja Award 
endowed by the Federal Ministry of Education and Research 
as well as funds from the Swedish Research Council 
(Grant nr. 2015-00415\_3) and 
Marie Sklodowska Curie Actions (Cofund Project INCA 600398).
PSB acknowledges support from the Royal Swedish Academy of Sciences, 
the Wenner-Gren Foundation, Goran Gustafssons Stiftelse and 
the Swedish Research Council. 
For much of this work PSB was a Royal Swedish Academy of 
Sciences Research Fellow supported by a grant from the 
Knut and Alice Wallenberg Foundation. 
PSB is presently partially supported by the project grant 
“The New Milky Way” from the Knut and Alice Wallenberg Foundation. 
Funding for the Stellar Astrophysics Centre is provided by 
The Danish National Research Foundation (Grant agreement no.: DNRF106).”
This research was undertaken with the 
assistance of resources from the 
National Computational Infrastructure (NCI),
which is supported by the Australian Government.

\appendix


\section{Line lists}
\label{linelist}

\begin{table*}
\begin{center}
\caption{\FeI lines analysed in the various benchmark stars. }
\label{linelisti}
\begin{tabular}{c l l c c c c c c c c}
\hline
\multirow{2}{*}{$\lambda_{\mathrm{Air}} / \mathrm{nm}$} &
\multirow{2}{*}{Lower level} &
\multirow{2}{*}{Upper level} &
\multirow{2}{*}{$J_{\mathrm{low}}$} &
\multirow{2}{*}{$J_{\mathrm{up}}$} &
\multirow{2}{*}{$E_{\mathrm{low}} / \mathrm{eV}$} &
\multirow{2}{*}{$\log{g\,f}$} &
\multicolumn{4}{c}{$\text{Equivalent width} / \text{pm}$} \\
& & & & & & &
HD84937 &
HD122563 &
HD140283 &
G64-12 \\
\hline
\hline
344.52 &
3d$^{7}$.($^{4}$P).4s     a$\,^{5}$P  &
3d$^{7}$.($^{4}$P).4p     u$\,^{5}$D  &
   2.0 &
   3.0 &
  2.1979 &
  -0.535$^g$ &
 &
 &
 &
 0.79 \\
347.67 &
3d$^{6}$.($^{5}$D).4s$^{2}$    a$\,^{5}$D  &
3d$^{6}$.($^{5}$D).4s.4p  z$\,^{5}$P  &
   0.0 &
   1.0 &
  0.1213 &
  -1.506$^g$ &
 &
 &
 &
 3.62 \\
386.55 &
3d$^{7}$.($^{4}$F).4s     a$\,^{5}$F  &
3d$^{7}$.($^{4}$F).4p     y$\,^{5}$D  &
   1.0 &
   1.0 &
  1.0111 &
  -0.950$^g$ &
 &
 &
 &
 3.41 \\
395.00 &
3d$^{7}$.($^{4}$P).4s     a$\,^{5}$P  &
3d$^{6}$.($^{3}$P).4s.4p  x$\,^{5}$P  &
   3.0 &
   2.0 &
  2.1759 &
  -1.251$^g$ &
 2.17 &
 &
 &
 \\
400.17 &
3d$^{7}$.($^{4}$P).4s     a$\,^{5}$P  &
3d$^{6}$.($^{3}$P).4s.4p  x$\,^{5}$P  &
   3.0 &
   3.0 &
  2.1759 &
  -1.901$^g$ &
 &
 &
 0.93 &
 \\
400.73 &
3d$^{7}$.($^{2}$G).4s     a$\,^{3}$G  &
3d$^{6}$.($^{3}$F).4s.4p  x$\,^{3}$F  &
   3.0 &
   2.0 &
  2.7586 &
  -1.276$^g$ &
 0.77 &
 &
 0.78 &
 \\
400.97 &
3d$^{7}$.($^{4}$P).4s     a$\,^{5}$P  &
3d$^{6}$.($^{3}$P).4s.4p  x$\,^{5}$P  &
   1.0 &
   2.0 &
  2.2227 &
  -1.252$^g$ &
 2.23 &
 &
 2.68 &
 \\
402.19 &
3d$^{7}$.($^{2}$G).4s     a$\,^{3}$G  &
3d$^{6}$.($^{3}$H).4s.4p  z$\,^{3}$H  &
   3.0 &
   4.0 &
  2.7586 &
  -0.729$^g$ &
 2.35 &
 &
 &
 \\
404.46 &
3d$^{7}$.($^{4}$P).4s     b$\,^{3}$P  &
3d$^{7}$.($^{4}$P).4p     y$\,^{3}$S  &
   2.0 &
   1.0 &
  2.8316 &
  -1.221$^g$ &
 0.92 &
 &
 1.06 &
 \\
406.24 &
3d$^{7}$.($^{4}$P).4s     b$\,^{3}$P  &
3d$^{7}$.($^{4}$P).4p     y$\,^{3}$S  &
   1.0 &
   1.0 &
  2.8450 &
  -0.862$^g$ &
 1.59 &
 &
 1.74 &
 \\
413.99 &
3d$^{7}$.($^{4}$F).4s     a$\,^{5}$F  &
3d$^{6}$.($^{5}$D).4s.4p  z$\,^{3}$F  &
   2.0 &
   2.0 &
  0.9901 &
  -3.514$^g$ &
 &
 3.64 &
 &
 \\
414.77 &
3d$^{7}$.($^{4}$F).4s     a$\,^{3}$F  &
3d$^{7}$.($^{4}$F).4p     z$\,^{3}$G  &
   4.0 &
   3.0 &
  1.4849 &
  -2.071$^g$ &
 1.74 &
 &
 &
 \\
415.88 &
3d$^{6}$.($^{5}$D).4s.4p  z$\,^{5}$F  &
3d$^{6}$.($^{5}$D).4s.4d  f$\,^{5}$F  &
   1.0 &
   2.0 &
  3.4302 &
  -0.700$^h$ &
 0.86 &
 &
 &
 \\
418.49 &
3d$^{7}$.($^{4}$P).4s     b$\,^{3}$P  &
3d$^{6}$.($^{3}$P).4s.4p  y$\,^{3}$P  &
   2.0 &
   2.0 &
  2.8316 &
  -0.869$^g$ &
 1.61 &
 &
 1.81 &
 \\
419.62 &
3d$^{6}$.($^{5}$D).4s.4p  z$\,^{5}$F  &
3d$^{6}$.($^{5}$D).4s.4d  e$\,^{5}$G  &
   3.0 &
   3.0 &
  3.3965 &
  -0.696$^g$ &
 0.86 &
 &
 0.94 &
 \\
420.20 &
3d$^{7}$.($^{4}$F).4s     a$\,^{3}$F  &
3d$^{7}$.($^{4}$F).4p     z$\,^{3}$G  &
   4.0 &
   4.0 &
  1.4849 &
  -0.689$^g$ &
 &
 &
 &
 2.82 \\
421.62 &
3d$^{6}$.($^{5}$D).4s$^{2}$    a$\,^{5}$D  &
3d$^{6}$.($^{5}$D).4s.4p  z$\,^{7}$P  &
   4.0 &
   4.0 &
  0.0000 &
  -3.357$^g$ &
 2.16 &
 &
 3.69 &
 \\
421.94 &
3d$^{7}$.($^{2}$H).4s     a$\,^{1}$H  &
3d$^{7}$.($^{2}$H).4p     y3I  &
   5.0 &
   6.0 &
  3.5732 &
   0.000$^g$ &
 2.56 &
 &
 2.48 &
 \\
422.22 &
3d$^{6}$.($^{5}$D).4s.4p  z$\,^{7}$D  &
3d$^{6}$.($^{5}$D).4s.5s  e$\,^{7}$D  &
   3.0 &
   3.0 &
  2.4496 &
  -0.914$^g$ &
 2.61 &
 &
 2.99 &
 \\
423.59 &
3d$^{6}$.($^{5}$D).4s.4p  z$\,^{7}$D  &
3d$^{6}$.($^{5}$D).4s.5s  e$\,^{7}$D  &
   4.0 &
   4.0 &
  2.4254 &
  -0.340$^i$ &
 &
 &
 &
 1.27 \\
423.88 &
3d$^{6}$.($^{5}$D).4s.4p  z$\,^{5}$F  &
3d$^{6}$.($^{5}$D).4s.4d  e$\,^{5}$G  &
   3.0 &
   4.0 &
  3.3965 &
  -0.233$^g$ &
 2.14 &
 &
 2.15 &
 \\
425.01 &
3d$^{6}$.($^{5}$D).4s.4p  z$\,^{7}$D  &
3d$^{6}$.($^{5}$D).4s.5s  e$\,^{7}$D  &
   2.0 &
   3.0 &
  2.4688 &
  -0.380$^g$ &
 4.97 &
 &
 &
 0.98 \\
426.05 &
3d$^{6}$.($^{5}$D).4s.4p  z$\,^{7}$D  &
3d$^{6}$.($^{5}$D).4s.5s  e$\,^{7}$D  &
   5.0 &
   5.0 &
  2.3992 &
   0.077$^g$ &
 &
 &
 &
 2.49 \\
428.24 &
3d$^{7}$.($^{4}$P).4s     a$\,^{5}$P  &
3d$^{6}$.($^{3}$P).4s.4p  z$\,^{5}$S  &
   3.0 &
   2.0 &
  2.1759 &
  -0.779$^g$ &
 4.27 &
 &
 4.81 &
 0.71 \\
437.59 &
3d$^{6}$.($^{5}$D).4s$^{2}$    a$\,^{5}$D  &
3d$^{6}$.($^{5}$D).4s.4p  z$\,^{7}$F  &
   4.0 &
   5.0 &
  0.0000 &
  -3.005$^g$ &
 3.72 &
 &
 &
 \\
440.48 &
3d$^{7}$.($^{4}$F).4s     a$\,^{3}$F  &
3d$^{7}$.($^{4}$F).4p     z$\,^{5}$G  &
   3.0 &
   4.0 &
  1.5574 &
  -0.147$^g$ &
 &
 &
 &
 5.26 \\
440.84 &
3d$^{7}$.($^{4}$P).4s     a$\,^{5}$P  &
3d$^{6}$.($^{5}$D).4s.4p  x$\,^{5}$D  &
   2.0 &
   1.0 &
  2.1979 &
  -1.775$^g$ &
 1.07 &
 &
 &
 \\
442.26 &
3d$^{7}$.($^{4}$P).4s     b$\,^{3}$P  &
3d$^{6}$.($^{3}$P).4s.4p  x$\,^{3}$D  &
   1.0 &
   1.0 &
  2.8450 &
  -1.115$^g$ &
 1.14 &
 &
 1.25 &
 \\
445.44 &
3d$^{7}$.($^{4}$P).4s     b$\,^{3}$P  &
3d$^{6}$.($^{3}$P).4s.4p  x$\,^{3}$D  &
   2.0 &
   2.0 &
  2.8316 &
  -1.298$^g$ &
 0.73 &
 &
 &
 \\
446.66 &
3d$^{7}$.($^{4}$P).4s     b$\,^{3}$P  &
3d$^{6}$.($^{3}$P).4s.4p  x$\,^{3}$D  &
   2.0 &
   3.0 &
  2.8316 &
  -0.600$^g$ &
 2.64 &
 &
 &
 \\
448.42 &
3d$^{6}$.($^{5}$D).4s.4p  z$\,^{5}$P  &
3d$^{6}$.($^{5}$D).4s.5s  g$\,^{5}$D  &
   3.0 &
   4.0 &
  3.6025 &
  -0.640$^h$ &
 0.61 &
 1.83 &
 &
 \\
449.46 &
3d$^{7}$.($^{4}$P).4s     a$\,^{5}$P  &
3d$^{6}$.($^{5}$D).4s.4p  x$\,^{5}$D  &
   2.0 &
   3.0 &
  2.1979 &
  -1.143$^g$ &
 2.86 &
 &
 3.63 &
 \\
452.86 &
3d$^{7}$.($^{4}$P).4s     a$\,^{5}$P  &
3d$^{6}$.($^{5}$D).4s.4p  x$\,^{5}$D  &
   3.0 &
   4.0 &
  2.1759 &
  -0.887$^g$ &
 &
 &
 &
 0.72 \\
453.12 &
3d$^{7}$.($^{4}$F).4s     a$\,^{3}$F  &
3d$^{7}$.($^{4}$F).4p     y$\,^{5}$F  &
   4.0 &
   4.0 &
  1.4849 &
  -2.101$^g$ &
 1.71 &
 &
 2.33 &
 \\
454.79 &
3d$^{7}$.(a$\,^{2}$D).4s    a$\,^{1}$D  &
3d$^{7}$.($^{2}$G).4p     z$\,^{1}$F  &
   2.0 &
   3.0 &
  3.5465 &
  -1.012$^g$ &
 &
 1.40 &
 &
 \\
461.93 &
3d$^{6}$.($^{5}$D).4s.4p  z$\,^{5}$P  &
3d$^{6}$.($^{5}$D).4s.4d  f$\,^{5}$D  &
   3.0 &
   2.0 &
  3.6025 &
  -1.060$^h$ &
 &
 0.91 &
 &
 \\
463.01 &
3d$^{6}$.($^{3}$P).4s$^{2}$    a$\,^{3}$P  &
3d$^{6}$.($^{5}$D).4s.4p  x$\,^{5}$D  &
   2.0 &
   3.0 &
  2.2786 &
  -2.587$^g$ &
 &
 1.50 &
 &
 \\
467.89 &
3d$^{6}$.($^{5}$D).4s.4p  z$\,^{5}$P  &
3d$^{6}$.($^{5}$D).4s.4d  f$\,^{5}$D  &
   3.0 &
   4.0 &
  3.6025 &
  -0.680$^h$ &
 &
 1.95 &
 &
 \\
473.68 &
3d$^{6}$.($^{5}$D).4s.4p  z$\,^{5}$D  &
3d$^{7}$.($^{4}$F).5s     e$\,^{5}$F  &
   4.0 &
   5.0 &
  3.2112 &
  -0.670$^h$ &
 1.31 &
 &
 1.44 &
 \\
474.15 &
3d$^{7}$.($^{4}$P).4s     b$\,^{3}$P  &
3d$^{6}$.($^{3}$P).4s.4p  w$\,^{5}$D  &
   2.0 &
   3.0 &
  2.8316 &
  -1.764$^g$ &
 &
 1.07 &
 &
 \\
478.88 &
3d$^{7}$.($^{2}$H).4s     b$\,^{3}$H  &
3d$^{6}$.($^{3}$H).4s.4p  z$\,^{3}$H  &
   6.0 &
   6.0 &
  3.2367 &
  -1.763$^g$ &
 &
 0.57 &
 &
 \\
488.21 &
3d$^{6}$.($^{5}$D).4s.4p  z$\,^{5}$F  &
3d$^{7}$.($^{4}$F).5s     e$\,^{5}$F  &
   2.0 &
   2.0 &
  3.4170 &
  -1.480$^h$ &
 &
 0.72 &
 &
 \\
491.90 &
3d$^{6}$.($^{5}$D).4s.4p  z$\,^{7}$F  &
3d$^{6}$.($^{5}$D).4s.5s  e$\,^{7}$D  &
   3.0 &
   3.0 &
  2.8654 &
  -0.342$^g$ &
 3.74 &
 &
 3.99 &
 \\
492.05 &
3d$^{6}$.($^{5}$D).4s.4p  z$\,^{7}$F  &
3d$^{6}$.($^{5}$D).4s.5s  e$\,^{7}$D  &
   5.0 &
   4.0 &
  2.8325 &
   0.070$^i$ &
 &
 &
 6.14 &
 1.38 \\
493.88 &
3d$^{6}$.($^{5}$D).4s.4p  z$\,^{7}$F  &
3d$^{6}$.($^{5}$D).4s.5s  e$\,^{7}$D  &
   2.0 &
   3.0 &
  2.8755 &
  -1.077$^g$ &
 1.02 &
 4.29 &
 &
 \\
494.64 &
3d$^{6}$.($^{5}$D).4s.4p  z$\,^{5}$F  &
3d$^{7}$.($^{4}$F).5s     e$\,^{5}$F  &
   4.0 &
   4.0 &
  3.3683 &
  -1.110$^i$ &
 &
 1.75 &
 &
 \\
495.01 &
3d$^{6}$.($^{5}$D).4s.4p  z$\,^{5}$F  &
3d$^{7}$.($^{4}$F).5s     e$\,^{5}$F  &
   2.0 &
   3.0 &
  3.4170 &
  -1.500$^h$ &
 &
 0.67 &
 &
 \\
496.61 &
3d$^{6}$.($^{5}$D).4s.4p  z$\,^{5}$F  &
3d$^{7}$.($^{4}$F).5s     e$\,^{5}$F  &
   5.0 &
   5.0 &
  3.3320 &
  -0.790$^h$ &
 0.88 &
 &
 &
 \\
497.31 &
3d$^{6}$.($^{5}$D).4s.4p  z$\,^{3}$D  &
3d$^{6}$.($^{5}$D).4s.5s  e$\,^{3}$D  &
   1.0 &
   1.0 &
  3.9597 &
  -0.690$^i$ &
 &
 0.90 &
 &
 \\
500.19 &
3d$^{6}$.($^{5}$D).4s.4p  z$\,^{3}$F  &
3d$^{6}$.($^{5}$D).4s.5s  e$\,^{3}$D  &
   4.0 &
   3.0 &
  3.8816 &
  -0.010$^i$ &
 &
 &
 1.38 &
 \\
500.61 &
3d$^{6}$.($^{5}$D).4s.4p  z$\,^{7}$F  &
3d$^{6}$.($^{5}$D).4s.5s  e$\,^{7}$D  &
   5.0 &
   5.0 &
  2.8325 &
  -0.615$^g$ &
 2.60 &
 &
 2.89 &
 \\
501.21 &
3d$^{7}$.($^{4}$F).4s     a$\,^{5}$F  &
3d$^{6}$.($^{5}$D).4s.4p  z$\,^{5}$F  &
   5.0 &
   5.0 &
  0.8590 &
  -2.642$^b$ &
 &
 &
 3.36 &
 \\
501.49 &
3d$^{6}$.($^{5}$D).4s.4p  z$\,^{3}$F  &
3d$^{6}$.($^{5}$D).4s.5s  e$\,^{3}$D  &
   3.0 &
   2.0 &
  3.9433 &
  -0.180$^h$ &
 0.96 &
 2.27 &
 &
 \\
502.22 &
3d$^{6}$.($^{5}$D).4s.4p  z$\,^{3}$F  &
3d$^{6}$.($^{5}$D).4s.5s  e$\,^{3}$D  &
   2.0 &
   1.0 &
  3.9841 &
  -0.330$^i$ &
 &
 1.40 &
 &
 \\
504.98 &
3d$^{6}$.($^{3}$P).4s$^{2}$    a$\,^{3}$P  &
3d$^{7}$.($^{4}$F).4p     y$\,^{3}$D  &
   2.0 &
   3.0 &
  2.2786 &
  -1.355$^g$ &
 1.80 &
 &
 2.25 &
 \\
505.16 &
3d$^{7}$.($^{4}$F).4s     a$\,^{5}$F  &
3d$^{6}$.($^{5}$D).4s.4p  z$\,^{5}$F  &
   4.0 &
   4.0 &
  0.9146 &
  -2.795$^b$ &
 &
 &
 2.39 &
 \\
506.88 &
3d$^{6}$.($^{5}$D).4s.4p  z$\,^{7}$P  &
3d$^{6}$.($^{5}$D).4s.5s  e$\,^{7}$D  &
   4.0 &
   3.0 &
  2.9398 &
  -1.041$^g$ &
 1.06 &
 4.07 &
 &
 \\
513.37 &
3d$^{7}$.($^{4}$F).4p     y$\,^{5}$F  &
3d$^{7}$.($^{4}$F).4d     f$\,^{5}$G  &
   5.0 &
   6.0 &
  4.1777 &
   0.360$^h$ &
 1.84 &
 &
 1.59 &
 \\
\hline
\hline
\end{tabular}
\end{center}
\end{table*}
 
\begin{table*}
\begin{center}
\contcaption{\FeI lines analysed in the various benchmark stars. }
\label{linelisti:continued}
\begin{tabular}{c l l c c c c c c c c}
\hline
\multirow{2}{*}{$\lambda_{\mathrm{Air}} / \mathrm{nm}$} &
\multirow{2}{*}{Lower level} &
\multirow{2}{*}{Upper level} &
\multirow{2}{*}{$J_{\mathrm{low}}$} &
\multirow{2}{*}{$J_{\mathrm{up}}$} &
\multirow{2}{*}{$E_{\mathrm{low}} / \mathrm{eV}$} &
\multirow{2}{*}{$\log{g\,f}$} &
\multicolumn{4}{c}{$\text{Equivalent width} / \text{pm}$} \\
& & & & & & &
HD84937 &
HD122563 &
HD140283 &
G64-12 \\
\hline
\hline
517.16 &
3d$^{7}$.($^{4}$F).4s     a$\,^{3}$F  &
3d$^{6}$.($^{5}$D).4s.4p  z$\,^{3}$F  &
   4.0 &
   4.0 &
  1.4849 &
  -1.721$^g$ &
 3.35 &
 &
 4.28 &
 \\
519.14 &
3d$^{6}$.($^{5}$D).4s.4p  z$\,^{7}$P  &
3d$^{6}$.($^{5}$D).4s.5s  e$\,^{7}$D  &
   2.0 &
   1.0 &
  3.0385 &
  -0.551$^g$ &
 &
 6.09 &
 2.28 &
 \\
519.23 &
3d$^{6}$.($^{5}$D).4s.4p  z$\,^{7}$P  &
3d$^{6}$.($^{5}$D).4s.5s  e$\,^{7}$D  &
   3.0 &
   3.0 &
  2.9980 &
  -0.421$^g$ &
 2.65 &
 &
 3.02 &
 \\
519.49 &
3d$^{7}$.($^{4}$F).4s     a$\,^{3}$F  &
3d$^{6}$.($^{5}$D).4s.4p  z$\,^{3}$F  &
   3.0 &
   3.0 &
  1.5574 &
  -2.021$^g$ &
 1.84 &
 &
 2.47 &
 \\
519.87 &
3d$^{7}$.($^{4}$P).4s     a$\,^{5}$P  &
3d$^{6}$.($^{5}$D).4s.4p  y$\,^{5}$P  &
   1.0 &
   2.0 &
  2.2227 &
  -2.135$^c$ &
 &
 3.87 &
 &
 \\
521.52 &
3d$^{6}$.($^{5}$D).4s.4p  z$\,^{5}$D  &
3d$^{6}$.($^{5}$D).4s.5s  e$\,^{5}$D  &
   2.0 &
   1.0 &
  3.2657 &
  -0.860$^h$ &
 &
 2.81 &
 &
 \\
521.63 &
3d$^{7}$.($^{4}$F).4s     a$\,^{3}$F  &
3d$^{6}$.($^{5}$D).4s.4p  z$\,^{3}$F  &
   2.0 &
   2.0 &
  1.6079 &
  -2.082$^g$ &
 1.37 &
 &
 2.10 &
 \\
522.55 &
3d$^{6}$.($^{5}$D).4s$^{2}$    a$\,^{5}$D  &
3d$^{6}$.($^{5}$D).4s.4p  z$\,^{7}$D  &
   1.0 &
   1.0 &
  0.1101 &
  -4.789$^a$ &
 &
 4.78 &
 &
 \\
523.29 &
3d$^{6}$.($^{5}$D).4s.4p  z$\,^{7}$P  &
3d$^{6}$.($^{5}$D).4s.5s  e$\,^{7}$D  &
   4.0 &
   5.0 &
  2.9398 &
  -0.057$^g$ &
 &
 &
 &
 0.75 \\
524.71 &
3d$^{6}$.($^{5}$D).4s$^{2}$    a$\,^{5}$D  &
3d$^{6}$.($^{5}$D).4s.4p  z$\,^{7}$D  &
   2.0 &
   3.0 &
  0.0873 &
  -4.946$^a$ &
 &
 3.91 &
 &
 \\
525.50 &
3d$^{6}$.($^{5}$D).4s$^{2}$    a$\,^{5}$D  &
3d$^{6}$.($^{5}$D).4s.4p  z$\,^{7}$D  &
   1.0 &
   2.0 &
  0.1101 &
  -4.764$^a$ &
 &
 5.28 &
 &
 \\
526.33 &
3d$^{6}$.($^{5}$D).4s.4p  z$\,^{5}$D  &
3d$^{6}$.($^{5}$D).4s.5s  e$\,^{5}$D  &
   2.0 &
   2.0 &
  3.2657 &
  -0.870$^h$ &
 0.74 &
 2.94 &
 &
 \\
528.18 &
3d$^{6}$.($^{5}$D).4s.4p  z$\,^{7}$P  &
3d$^{6}$.($^{5}$D).4s.5s  e$\,^{7}$D  &
   2.0 &
   3.0 &
  3.0385 &
  -0.833$^g$ &
 1.27 &
 4.63 &
 1.48 &
 \\
528.36 &
3d$^{6}$.($^{5}$D).4s.4p  z$\,^{5}$D  &
3d$^{6}$.($^{5}$D).4s.5s  e$\,^{5}$D  &
   3.0 &
   3.0 &
  3.2410 &
  -0.450$^h$ &
 &
 &
 2.06 &
 \\
530.23 &
3d$^{6}$.($^{5}$D).4s.4p  z$\,^{5}$D  &
3d$^{6}$.($^{5}$D).4s.5s  e$\,^{5}$D  &
   1.0 &
   2.0 &
  3.2830 &
  -0.730$^h$ &
 &
 3.65 &
 1.10 &
 \\
530.74 &
3d$^{7}$.($^{4}$F).4s     a$\,^{3}$F  &
3d$^{6}$.($^{5}$D).4s.4p  z$\,^{3}$F  &
   2.0 &
   3.0 &
  1.6079 &
  -2.912$^g$ &
 &
 3.59 &
 &
 \\
532.42 &
3d$^{6}$.($^{5}$D).4s.4p  z$\,^{5}$D  &
3d$^{6}$.($^{5}$D).4s.5s  e$\,^{5}$D  &
   4.0 &
   4.0 &
  3.2112 &
  -0.110$^i$ &
 3.24 &
 &
 3.47 &
 \\
533.99 &
3d$^{6}$.($^{5}$D).4s.4p  z$\,^{5}$D  &
3d$^{6}$.($^{5}$D).4s.5s  e$\,^{5}$D  &
   2.0 &
   3.0 &
  3.2657 &
  -0.630$^h$ &
 1.17 &
 4.19 &
 1.34 &
 \\
534.10 &
3d$^{7}$.($^{4}$F).4s     a$\,^{3}$F  &
3d$^{6}$.($^{5}$D).4s.4p  z$\,^{3}$D  &
   2.0 &
   2.0 &
  1.6079 &
  -1.953$^g$ &
 2.15 &
 &
 &
 \\
536.49 &
3d$^{7}$.($^{4}$F).4p     z$\,^{5}$G  &
3d$^{7}$.($^{4}$F).4d     e$\,^{5}$H  &
   2.0 &
   3.0 &
  4.4456 &
   0.228$^g$ &
 1.04 &
 &
 &
 \\
536.75 &
3d$^{7}$.($^{4}$F).4p     z$\,^{5}$G  &
3d$^{7}$.($^{4}$F).4d     e$\,^{5}$H  &
   3.0 &
   4.0 &
  4.4153 &
   0.443$^g$ &
 1.33 &
 &
 1.23 &
 \\
537.00 &
3d$^{7}$.($^{4}$F).4p     z$\,^{5}$G  &
3d$^{7}$.($^{4}$F).4d     e$\,^{5}$H  &
   4.0 &
   5.0 &
  4.3714 &
   0.536$^g$ &
 1.78 &
 2.97 &
 &
 \\
537.15 &
3d$^{7}$.($^{4}$F).4s     a$\,^{5}$F  &
3d$^{6}$.($^{5}$D).4s.4p  z$\,^{5}$D  &
   3.0 &
   2.0 &
  0.9582 &
  -1.645$^b$ &
 6.71 &
 &
 &
 \\
538.34 &
3d$^{7}$.($^{4}$F).4p     z$\,^{5}$G  &
3d$^{7}$.($^{4}$F).4d     e$\,^{5}$H  &
   5.0 &
   6.0 &
  4.3125 &
   0.645$^g$ &
 &
 3.73 &
 1.97 &
 \\
539.32 &
3d$^{6}$.($^{5}$D).4s.4p  z$\,^{5}$D  &
3d$^{6}$.($^{5}$D).4s.5s  e$\,^{5}$D  &
   3.0 &
   4.0 &
  3.2410 &
  -0.720$^i$ &
 &
 &
 1.23 &
 \\
539.71 &
3d$^{7}$.($^{4}$F).4s     a$\,^{5}$F  &
3d$^{6}$.($^{5}$D).4s.4p  z$\,^{5}$D  &
   4.0 &
   4.0 &
  0.9146 &
  -1.993$^b$ &
 5.02 &
 &
 6.58 &
 \\
541.09 &
3d$^{7}$.($^{4}$F).4p     z$\,^{3}$G  &
3d$^{7}$.($^{4}$F).4d     e$\,^{3}$H  &
   3.0 &
   4.0 &
  4.4733 &
   0.398$^g$ &
 1.17 &
 2.05 &
 &
 \\
541.52 &
3d$^{7}$.($^{4}$F).4p     z$\,^{3}$G  &
3d$^{7}$.($^{4}$F).4d     e$\,^{3}$H  &
   5.0 &
   6.0 &
  4.3865 &
   0.643$^g$ &
 2.01 &
 3.37 &
 1.66 &
 \\
542.97 &
3d$^{7}$.($^{4}$F).4s     a$\,^{5}$F  &
3d$^{6}$.($^{5}$D).4s.4p  z$\,^{5}$D  &
   3.0 &
   3.0 &
  0.9582 &
  -1.879$^b$ &
 5.62 &
 &
 &
 \\
543.45 &
3d$^{7}$.($^{4}$F).4s     a$\,^{5}$F  &
3d$^{6}$.($^{5}$D).4s.4p  z$\,^{5}$D  &
   1.0 &
   0.0 &
  1.0111 &
  -2.122$^b$ &
 &
 &
 5.37 &
 \\
544.69 &
3d$^{7}$.($^{4}$F).4s     a$\,^{5}$F  &
3d$^{6}$.($^{5}$D).4s.4p  z$\,^{5}$D  &
   2.0 &
   2.0 &
  0.9901 &
  -1.914$^g$ &
 &
 &
 &
 1.05 \\
549.75 &
3d$^{7}$.($^{4}$F).4s     a$\,^{5}$F  &
3d$^{6}$.($^{5}$D).4s.4p  z$\,^{5}$D  &
   1.0 &
   2.0 &
  1.0111 &
  -2.849$^b$ &
 &
 &
 1.93 &
 \\
550.68 &
3d$^{7}$.($^{4}$F).4s     a$\,^{5}$F  &
3d$^{6}$.($^{5}$D).4s.4p  z$\,^{5}$D  &
   2.0 &
   3.0 &
  0.9901 &
  -2.797$^b$ &
 1.43 &
 &
 &
 \\
556.96 &
3d$^{6}$.($^{5}$D).4s.4p  z$\,^{5}$F  &
3d$^{6}$.($^{5}$D).4s.5s  e$\,^{5}$D  &
   2.0 &
   1.0 &
  3.4170 &
  -0.520$^h$ &
 1.25 &
 3.82 &
 1.29 &
 \\
557.28 &
3d$^{6}$.($^{5}$D).4s.4p  z$\,^{5}$F  &
3d$^{6}$.($^{5}$D).4s.5s  e$\,^{5}$D  &
   3.0 &
   2.0 &
  3.3965 &
  -0.280$^h$ &
 2.09 &
 &
 1.99 &
 \\
558.68 &
3d$^{6}$.($^{5}$D).4s.4p  z$\,^{5}$F  &
3d$^{6}$.($^{5}$D).4s.5s  e$\,^{5}$D  &
   4.0 &
   3.0 &
  3.3683 &
  -0.110$^h$ &
 2.65 &
 6.31 &
 2.84 &
 \\
561.56 &
3d$^{6}$.($^{5}$D).4s.4p  z$\,^{5}$F  &
3d$^{6}$.($^{5}$D).4s.5s  e$\,^{5}$D  &
   5.0 &
   4.0 &
  3.3320 &
   0.040$^i$ &
 3.38 &
 &
 &
 \\
562.45 &
3d$^{6}$.($^{5}$D).4s.4p  z$\,^{5}$F  &
3d$^{6}$.($^{5}$D).4s.5s  e$\,^{5}$D  &
   2.0 &
   2.0 &
  3.4170 &
  -0.760$^h$ &
 &
 2.73 &
 &
 \\
576.30 &
3d$^{6}$.($^{5}$D).4s.4p  z$\,^{3}$P  &
3d$^{6}$.($^{5}$D).4s.5s  e$\,^{3}$D  &
   2.0 &
   3.0 &
  4.2089 &
  -0.360$^i$ &
 &
 1.03 &
 &
 \\
595.67 &
3d$^{7}$.($^{4}$F).4s     a$\,^{5}$F  &
3d$^{6}$.($^{5}$D).4s.4p  z$\,^{7}$P  &
   5.0 &
   4.0 &
  0.8590 &
  -4.608$^e$ &
 &
 1.27 &
 &
 \\
619.16 &
3d$^{6}$.($^{3}$H).4s$^{2}$    a$\,^{3}$H  &
3d$^{7}$.($^{4}$F).4p     z$\,^{3}$G  &
   5.0 &
   4.0 &
  2.4327 &
  -1.416$^g$ &
 1.33 &
 &
 1.80 &
 \\
621.34 &
3d$^{7}$.($^{4}$P).4s     a$\,^{5}$P  &
3d$^{7}$.($^{4}$F).4p     y$\,^{5}$D  &
   1.0 &
   1.0 &
  2.2227 &
  -2.481$^g$ &
 &
 2.20 &
 &
 \\
621.93 &
3d$^{7}$.($^{4}$P).4s     a$\,^{5}$P  &
3d$^{7}$.($^{4}$F).4p     y$\,^{5}$D  &
   2.0 &
   2.0 &
  2.1979 &
  -2.433$^c$ &
 &
 2.76 &
 &
 \\
623.07 &
3d$^{6}$.($^{3}$F).4s$^{2}$    b$\,^{3}$F  &
3d$^{7}$.($^{4}$F).4p     y$\,^{3}$F  &
   4.0 &
   4.0 &
  2.5592 &
  -1.281$^d$ &
 1.70 &
 6.89 &
 1.97 &
 \\
625.43 &
3d$^{6}$.($^{3}$P).4s$^{2}$    a$\,^{3}$P  &
3d$^{6}$.($^{5}$D).4s.4p  z$\,^{3}$P  &
   2.0 &
   1.0 &
  2.2786 &
  -2.443$^f$ &
 &
 2.62 &
 &
 \\
626.51 &
3d$^{7}$.($^{4}$P).4s     a$\,^{5}$P  &
3d$^{7}$.($^{4}$F).4p     y$\,^{5}$D  &
   3.0 &
   3.0 &
  2.1759 &
  -2.550$^c$ &
 &
 2.60 &
 &
 \\
629.78 &
3d$^{7}$.($^{4}$P).4s     a$\,^{5}$P  &
3d$^{7}$.($^{4}$F).4p     y$\,^{5}$D  &
   1.0 &
   2.0 &
  2.2227 &
  -2.740$^c$ &
 &
 1.49 &
 &
 \\
633.53 &
3d$^{7}$.($^{4}$P).4s     a$\,^{5}$P  &
3d$^{7}$.($^{4}$F).4p     y$\,^{5}$D  &
   2.0 &
   3.0 &
  2.1979 &
  -2.177$^g$ &
 &
 4.02 &
 &
 \\
633.68 &
3d$^{6}$.($^{5}$D).4s.4p  z$\,^{5}$P  &
3d$^{6}$.($^{5}$D).4s.5s  e$\,^{5}$D  &
   1.0 &
   1.0 &
  3.6864 &
  -0.850$^h$ &
 &
 1.32 &
 &
 \\
635.87 &
3d$^{7}$.($^{4}$F).4s     a$\,^{5}$F  &
3d$^{6}$.($^{5}$D).4s.4p  z$\,^{7}$F  &
   5.0 &
   6.0 &
  0.8590 &
  -4.468$^b$ &
 &
 1.94 &
 &
 \\
639.36 &
3d$^{6}$.($^{3}$H).4s$^{2}$    a$\,^{3}$H  &
3d$^{7}$.($^{4}$F).4p     z$\,^{5}$G  &
   5.0 &
   4.0 &
  2.4327 &
  -1.432$^f$ &
 &
 6.24 &
 &
 \\
640.00 &
3d$^{6}$.($^{5}$D).4s.4p  z$\,^{5}$P  &
3d$^{6}$.($^{5}$D).4s.5s  e$\,^{5}$D  &
   3.0 &
   4.0 &
  3.6025 &
  -0.270$^i$ &
 1.51 &
 &
 1.58 &
 \\
641.16 &
3d$^{6}$.($^{5}$D).4s.4p  z$\,^{5}$P  &
3d$^{6}$.($^{5}$D).4s.5s  e$\,^{5}$D  &
   2.0 &
   3.0 &
  3.6537 &
  -0.590$^h$ &
 &
 2.46 &
 &
 \\
642.14 &
3d$^{6}$.($^{3}$P).4s$^{2}$    a$\,^{3}$P  &
3d$^{6}$.($^{5}$D).4s.4p  z$\,^{3}$P  &
   2.0 &
   2.0 &
  2.2786 &
  -2.027$^c$ &
 &
 4.76 &
 &
 \\
643.08 &
3d$^{7}$.($^{4}$P).4s     a$\,^{5}$P  &
3d$^{7}$.($^{4}$F).4p     y$\,^{5}$D  &
   3.0 &
   4.0 &
  2.1759 &
  -2.006$^c$ &
 &
 5.78 &
 &
 \\
649.50 &
3d$^{6}$.($^{3}$H).4s$^{2}$    a$\,^{3}$H  &
3d$^{7}$.($^{4}$F).4p     z$\,^{5}$G  &
   6.0 &
   5.0 &
  2.4041 &
  -1.273$^c$ &
 &
 &
 2.53 &
 \\
654.62 &
3d$^{7}$.($^{2}$G).4s     a$\,^{3}$G  &
3d$^{7}$.($^{4}$F).4p     y$\,^{3}$F  &
   3.0 &
   2.0 &
  2.7586 &
  -1.536$^g$ &
 &
 3.34 &
 &
 \\
659.29 &
3d$^{7}$.($^{2}$G).4s     a$\,^{3}$G  &
3d$^{7}$.($^{4}$F).4p     y$\,^{3}$F  &
   4.0 &
   3.0 &
  2.7275 &
  -1.473$^g$ &
 &
 3.84 &
 &
 \\
\hline
\hline
\end{tabular}
\end{center}
\end{table*}
 
\begin{table*}
\begin{center}
\contcaption{\FeI lines analysed in the various benchmark stars. }
\label{linelisti:continued}
\begin{tabular}{c l l c c c c c c c c}
\hline
\multirow{2}{*}{$\lambda_{\mathrm{Air}} / \mathrm{nm}$} &
\multirow{2}{*}{Lower level} &
\multirow{2}{*}{Upper level} &
\multirow{2}{*}{$J_{\mathrm{low}}$} &
\multirow{2}{*}{$J_{\mathrm{up}}$} &
\multirow{2}{*}{$E_{\mathrm{low}} / \mathrm{eV}$} &
\multirow{2}{*}{$\log{g\,f}$} &
\multicolumn{4}{c}{$\text{Equivalent width} / \text{pm}$} \\
& & & & & & &
HD84937 &
HD122563 &
HD140283 &
G64-12 \\
\hline
\hline
667.80 &
3d$^{7}$.($^{2}$G).4s     a$\,^{3}$G  &
3d$^{7}$.($^{4}$F).4p     y$\,^{3}$F  &
   5.0 &
   4.0 &
  2.6924 &
  -1.418$^g$ &
 &
 5.10 &
 &
 \\
675.01 &
3d$^{6}$.($^{3}$P).4s$^{2}$    a$\,^{3}$P  &
3d$^{6}$.($^{5}$D).4s.4p  z$\,^{3}$P  &
   1.0 &
   1.0 &
  2.4242 &
  -2.621$^c$ &
 &
 1.34 &
 &
 \\
\hline
\hline
\end{tabular}
\end{center}
\medskip
a: \citet{1979MNRAS.186..633B}; 
b: \citet{1979MNRAS.186..657B}; 
c: \citet{1982MNRAS.199...43B}; 
d: \citet{1982MNRAS.201..595B}; 
e: \citet{1986MNRAS.220..289B}; 
f: \citet{1991A&amp;A...248..315B}; 
g: \citet{1991JOSAB...8.1185O}; 
h: \citet{2014ApJS..215...23D}; 
i: \citet{2014MNRAS.441.3127R}.
\end{table*}

\begin{table*}
\begin{center}
\caption{\FeII lines analysed in the various benchmark stars. \markaschanged{The excitation energies are given relative to the ground state of \FeI,such that the ground state of \FeII~has $E_{\mathrm{low}}=7.9024\,\mathrm{eV}$.}}
\label{linelistii}
\begin{tabular}{c l l c c c c c c c c}
\hline
\multirow{2}{*}{$\lambda_{\mathrm{Air}} / \mathrm{nm}$} &
\multirow{2}{*}{Lower level} &
\multirow{2}{*}{Upper level} &
\multirow{2}{*}{$J_{\mathrm{low}}$} &
\multirow{2}{*}{$J_{\mathrm{up}}$} &
\multirow{2}{*}{$E_{\mathrm{low}} / \mathrm{eV}$} &
\multirow{2}{*}{$\log{g\,f}$} &
\multicolumn{4}{c}{$\text{Equivalent width} / \text{pm}$} \\
& & & & & & &
HD84937 &
HD122563 &
HD140283 &
G64-12 \\
\hline
\hline
417.89 &
3d$^{6}$.($^{3}$P).4s     b$\,^{4}$P  &
3d$^{6}$.($^{5}$D).4p     z$\,^{4}$F  &
   2.5 &
   3.5 &
 10.4850 &
  -2.510$^a$ &
 2.10 &
 &
 &
 \\
430.32 &
3d$^{6}$.($^{3}$P).4s     b$\,^{4}$P  &
3d$^{6}$.($^{5}$D).4p     z$\,^{4}$D  &
   1.5 &
   1.5 &
 10.6067 &
  -2.560$^a$ &
 1.77 &
 &
 &
 \\
438.54 &
3d$^{6}$.($^{3}$P).4s     b$\,^{4}$P  &
3d$^{6}$.($^{5}$D).4p     z$\,^{4}$D  &
   0.5 &
   0.5 &
 10.6808 &
  -2.660$^a$ &
 1.59 &
 &
 &
 \\
441.68 &
3d$^{6}$.($^{3}$P).4s     b$\,^{4}$P  &
3d$^{6}$.($^{5}$D).4p     z$\,^{4}$D  &
   0.5 &
   1.5 &
 10.6808 &
  -2.650$^a$ &
 1.35 &
 4.24 &
 1.21 &
 \\
449.14 &
3d$^{6}$.($^{3}$F).4s     b$\,^{4}$F  &
3d$^{6}$.($^{5}$D).4p     z$\,^{4}$F  &
   1.5 &
   1.5 &
 10.7579 &
  -2.710$^a$ &
 0.95 &
 3.43 &
 0.75 &
 \\
450.83 &
3d$^{6}$.($^{3}$F).4s     b$\,^{4}$F  &
3d$^{6}$.($^{5}$D).4p     z$\,^{4}$D  &
   1.5 &
   0.5 &
 10.7579 &
  -2.440$^a$ &
 1.92 &
 5.30 &
 1.70 &
 \\
451.53 &
3d$^{6}$.($^{3}$F).4s     b$\,^{4}$F  &
3d$^{6}$.($^{5}$D).4p     z$\,^{4}$F  &
   2.5 &
   2.5 &
 10.7465 &
  -2.600$^a$ &
 1.49 &
 4.69 &
 1.29 &
 \\
452.02 &
3d$^{6}$.($^{3}$F).4s     b$\,^{4}$F  &
3d$^{6}$.($^{5}$D).4p     z$\,^{4}$F  &
   4.5 &
   3.5 &
 10.7090 &
  -2.650$^a$ &
 1.40 &
 4.47 &
 1.16 &
 \\
454.15 &
3d$^{6}$.($^{3}$F).4s     b$\,^{4}$F  &
3d$^{6}$.($^{5}$D).4p     z$\,^{4}$D  &
   1.5 &
   1.5 &
 10.7579 &
  -2.980$^a$ &
 &
 2.50 &
 &
 \\
457.63 &
3d$^{6}$.($^{3}$F).4s     b$\,^{4}$F  &
3d$^{6}$.($^{5}$D).4p     z$\,^{4}$D  &
   2.5 &
   2.5 &
 10.7465 &
  -2.950$^a$ &
 0.70 &
 2.38 &
 &
 \\
458.38 &
3d$^{6}$.($^{3}$F).4s     b$\,^{4}$F  &
3d$^{6}$.($^{5}$D).4p     z$\,^{4}$D  &
   4.5 &
   3.5 &
 10.7090 &
  -1.930$^a$ &
 4.39 &
 &
 3.87 &
 0.71 \\
462.05 &
3d$^{6}$.($^{3}$F).4s     b$\,^{4}$F  &
3d$^{6}$.($^{5}$D).4p     z$\,^{4}$D  &
   3.5 &
   3.5 &
 10.7305 &
  -3.210$^a$ &
 &
 1.31 &
 &
 \\
466.68 &
3d$^{6}$.($^{3}$F).4s     b$\,^{4}$F  &
3d$^{6}$.($^{5}$D).4p     z$\,^{4}$F  &
   3.5 &
   4.5 &
 10.7305 &
  -3.280$^a$ &
 &
 1.28 &
 &
 \\
492.39 &
3d$^{5}$.($^{6}$S).4s$^{2}$    a$\,^{6}$S  &
3d$^{6}$.($^{5}$D).4p     z$\,^{6}$P  &
   2.5 &
   1.5 &
 10.7934 &
  -1.260$^a$ &
 &
 &
 5.92 &
 \\
501.84 &
3d$^{5}$.($^{6}$S).4s$^{2}$    a$\,^{6}$S  &
3d$^{6}$.($^{5}$D).4p     z$\,^{6}$P  &
   2.5 &
   2.5 &
 10.7934 &
  -1.100$^a$ &
 &
 &
 &
 1.86 \\
516.90 &
3d$^{5}$.($^{6}$S).4s$^{2}$    a$\,^{6}$S  &
3d$^{6}$.($^{5}$D).4p     z$\,^{6}$P  &
   2.5 &
   3.5 &
 10.7934 &
  -1.000$^a$ &
 &
 &
 &
 2.46 \\
519.76 &
3d$^{6}$.($^{3}$G).4s     a$\,^{4}$G  &
3d$^{6}$.($^{5}$D).4p     z$\,^{4}$F  &
   2.5 &
   1.5 &
 11.1328 &
  -2.220$^a$ &
 1.34 &
 3.92 &
 1.16 &
 \\
523.46 &
3d$^{6}$.($^{3}$G).4s     a$\,^{4}$G  &
3d$^{6}$.($^{5}$D).4p     z$\,^{4}$F  &
   3.5 &
   2.5 &
 11.1237 &
  -2.180$^a$ &
 1.68 &
 4.30 &
 1.45 &
 \\
527.60 &
3d$^{6}$.($^{3}$G).4s     a$\,^{4}$G  &
3d$^{6}$.($^{5}$D).4p     z$\,^{4}$F  &
   4.5 &
   3.5 &
 11.1018 &
  -2.010$^a$ &
 2.03 &
 &
 1.94 &
 \\
531.66 &
3d$^{6}$.($^{3}$G).4s     a$\,^{4}$G  &
3d$^{6}$.($^{5}$D).4p     z$\,^{4}$F  &
   5.5 &
   4.5 &
 11.0551 &
  -1.870$^a$ &
 3.43 &
 &
 2.65 &
 \\
536.29 &
3d$^{6}$.($^{3}$G).4s     a$\,^{4}$G  &
3d$^{6}$.($^{5}$D).4p     z$\,^{4}$D  &
   4.5 &
   3.5 &
 11.1018 &
  -2.570$^a$ &
 0.86 &
 2.83 &
 &
 \\
553.48 &
3d$^{6}$.($^{3}$H).4s     b$\,^{2}$H  &
3d$^{6}$.($^{5}$D).4p     z$\,^{4}$F  &
   5.5 &
   4.5 &
 11.1471 &
  -2.750$^a$ &
 &
 1.73 &
 &
 \\
645.64 &
3d$^{6}$.($^{3}$D).4s     b$\,^{4}$D  &
3d$^{6}$.($^{5}$D).4p     z$\,^{4}$P  &
   3.5 &
   2.5 &
 11.8058 &
  -2.050$^a$ &
 &
 1.45 &
 &
 \\
\hline
\hline
\end{tabular}
\end{center}
\medskip
a: \citet{2009A&amp;A...497..611M}.
\end{table*}

We list the lines used for the abundance analyses of the benchmark stars
in \tab{linelisti}~and \tab{linelistii}.

\bibliographystyle{mnras}
\bibliography{/Users/ama51/Documents/work/papers/allpapers/bibl.bib}

\label{lastpage}
\end{document}